\documentclass{aa}  
\usepackage{blindtext}
\usepackage{longtable}
\usepackage{graphicx}
\usepackage{subcaption}  
\usepackage{soul}

\usepackage{txfonts}

\usepackage{xcolor}
\begin{document}

\title{Emergence of magnetic flux sheets in the quiet Sun}
\subtitle{I. Statistical properties}

 \author{  
            S. M. Díaz-Castillo\inst{1,2}
            \and
            C. E. Fischer\inst{3,4}
            \and
            F. Moreno-Insertis\inst{5,6}
            \and
            S. L. Guglielmino\inst{7}
            \and
            R. Ishikawa\inst{8}
            \and
            S. Criscuoli\inst{4}
          }

   \institute{Institut f\"{u}r Sonnenphysik, 
             Georges-K\"ohler-Allee 401 A, Freiburg i.Br., Germany \\
              \email{smdiazcas@leibniz-kis.de}
         \and
            Institute of Physics, Faculty of Mathematics and Physics, University of Freiburg, Freiburg i.Br., Germany 
         \and 
            European Space Agency (ESA)
            European Space Astronomy Centre (ESAC), Madrid, Spain 
         \and
             National Solar Observatory, 
             3665 Discovery Dr., Boulder, CO 80303, USA 
        \and
             Instituto de Astrof\'{i}sica de Canarias, 
             Avda. V\'{i}a L\'{a}ctea S/N,
             38200 La Laguna, Tenerife, Spain
        \and
             Departamento de Astrof\'{i}sica,
             Universidad de La Laguna,
             38205 La Laguna, Tenerife, Spain   
        \and
            INAF - Osservatorio Astrofisico di Catania, 
            Via Santa Sofia 78, 95123 Catania, Italia
        \and
            National Astronomical Observatory of Japan, 
            2-21-1 Osawa, Mitaka, 
            Tokyo 181-8588, Japan      
           }

   \date{Received 05.12.2024; accepted: 26.01.2025}
   
     \abstract
   {Small-scale magnetic flux emergence in the quiet Sun is crucial for maintaining solar magnetic activity. On the smallest scales studied so far, namely within individual granules, two mechanisms have been identified: emergence in tiny magnetic loops and emergence in the form of magnetic flux sheets covering the granule. While there are abundant observations of tiny magnetic loops within granules, the evidence for the emergence of granule-covering magnetic sheets is much more limited.}
   {This work aims to statistically analyze magnetic flux sheets, quantify their frequency on the solar surface and their potential contribution to the solar magnetic budget in the photosphere, and investigate the plasma dynamics and granular-scale phenomena associated with their emergence.}
   {Using spectro-polarimetric datasets taken along the \ion{Fe}{I} 630.15 and 630.25\,nm photospheric lines from the solar optical telescope aboard the Hinode satellite and the \ion{Fe}{I} 630.15, 630.25\,nm and 617.3\,nm from the ground-based Swedish Solar telescope, we developed a two-step method to identify magnetic flux sheet emergence events, detecting magnetic flux patches based on the calculation of the transverse and longitudinal magnetic flux density and associating them with their host granules based on velocity field analysis.}
   {We identified 42 events of magnetic flux sheet emergence and characterized their magnetic properties and the associated plasma dynamics of their host granules. Our results align with numerical simulations, indicating a similar occurrence rate of approximately 0.3 events per day per Mm$^2$. We investigated the relationship between magnetic flux emergence and granular phenomena, finding that flux sheets often emerge in association with standard nascent granules, as well as exploding granules, or granules with granular lanes. In particular, we highlight the potential role of recycled magnetic flux from downflow regions in facilitating flux sheet emergence. Our analysis suggests that the magnetic flux sheet events could be considered part of the larger component of the distribution of small-scale magnetic flux that feeds the solar atmosphere in quiet Sun regions.}
   {}


   \maketitle
%

\section{Introduction} \label{sec:intro}

Estimates of the emergence rate of the global magnetic flux show a staggering amount of, in some cases, more than $10^{25}\,\rm{Mx}$ of magnetic flux emerging through the solar surface per day \citep{ThornthonParnell2011}. One important contribution to the global magnetic flux budget is the small-scale magnetic flux in the quiet Sun, on spatial scales ranging from 5" down to the current observational limit. It supplies between $10^{23}$ to $10^{24}\,\rm{Mx}$ of the total magnetic flux at anytime \citep{Simon2001, Hagenaar2003, Zhou2013, Gosic2015}, which is similar to the amount present in active regions during a solar maximum \citep{Jin2011}. Much of this magnetic flux is provided by network regions \citep[for a review, see][and references therein]{BellorRubio2019}, defined as regions of relatively strong and vertical magnetic fields occupying the outer boundaries of supergranular cells. \cite{Gosic2014} reported that 40\% of the magnetic flux emerged inside of the supergranular cells, in the internetwork regions \citep[see][and references therein]{BellorRubio2019}, ends up in the network, transferring flux at a rate of $1.5\times10^{24}\,\rm{Mx}$ per day. This highlights the importance of the extremely dynamic internetwork regions in supplying magnetic flux to the network.

In internetwork regions, the emergence of the magnetic flux mainly occurs in the form of small-scale magnetic loops within granules, which are advected toward the surface by the upward motion of the plasma. Spectro-polarimetric measurements of Hinode allowed researchers to trace small-scale magnetic loops from their initial appearance \citep{MartinezGonzalez2007, Centeno2007, Gomory2010, Ishikawa2010}. A linear polarization patch appears when the loop emerges. Simultaneously or shortly after, the footpoints are visible as longitudinal magnetic field patches of opposite polarity connected to the linear polarization patch. Finally, the footpoints move apart and are dragged to the intergranular lanes, while the apex disappears. Small-scale loops can rise and eventually reach chromospheric heights transferring energy into upper layers of the atmosphere \citep{MartinezGonzalez2009,Ishikawa2010, Smitha2017}. 

Another mode of flux emergence within granules, namely as magnetic flux sheets, was discovered in the framework of 3D magnetoconvection simulations by \cite{MorenoInsertis2018} in addition to the emergence of small-scale loops. Those authors found that these two emergence modes differ, first of all, in the fraction of granular surface they cover along their evolution, and in the granular evolutionary stage in which they appear. On the one hand, small-scale magnetic loops are concentrated, tube-like magnetic structures that emerge in well-formed granules. On the other hand, flux sheet emergence results from the stretching of a magnetic flux structure near the photosphere in expanding phases of the host granule, leading to an organised horizontal magnetic "mantle" as obtained in the simulation of \citet{MorenoInsertis2018} and also in the observations of \citet{Fischer2019}. In the latter paper, as the sheet expands towards the granule's edges, it eventually fragments, leaving only the footpoints, which can be observed as opposite-polarity patches in circular polarization maps. In fact, the magnetic flux sheet that emerged exhibited larger magnetic flux than the standard loop emergence and showed small clusters of mixed-polarity elements surrounding it, similar to those later analyzed by \cite{Gošić_2022}. At an even stronger emerged magnetic flux, the magnetic structure might interact with the convective plasma during emergence, shaping the flux appearance. In special cases like active regions, this interaction leads to elongated granulation, indicating the arrival of strong transverse magnetic flux \citep{Centeno2012, Centeno2017, Wang2020}. 

High-resolution observations of the flux emergence process at granular scales can help to validate the mechanisms of small-scale magnetic field generation identified in numerical simulations \citep{Rempel2023}. Statistical characterization of magnetic flux sheets is a way to shed light on how they emerge at the surface and interact with the photospheric flows, how large their contribution is to the energy budget of the solar atmosphere, and on the conditions under which they can reach chromospheric heights. This work aims to provide a statistical analysis of photospheric magnetic flux sheets and quantify the amount of magnetic flux they contribute at the photospheric level. We study the plasma dynamics underlying flux sheet emergence in the host granule and the relation with different stages and structural features of the granules such as exploding granules \citep{Rast1995, Palacios2012, Fischer2017} granules in their initial stages of evolution or nascent granules \citep{Bray1984, Stix2002, MorenoInsertis2018} and granular lanes \citep{Steiner2010, Fischer2020}.

\begin{figure} [h!]
	\centering
	\includegraphics[width=0.5\textwidth]{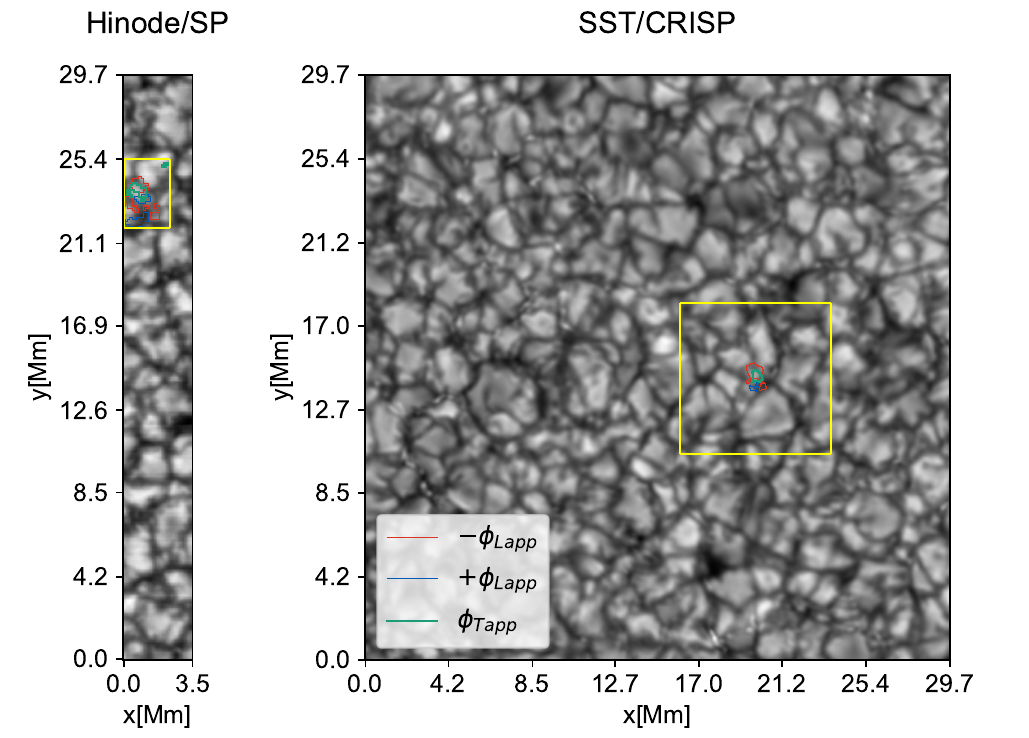}
	\caption{Left: Hinode SP full-sized and reconstructed image from slit data in the continuum of the \ion{Fe}{I} 630.2\,nm line pair with a field-of-view of $4\farcs5 \times 41\arcsec$ recorded on 2007-Oct-22 18:34UT. Right: Image of the CRISP instrument in the continuum of the \ion{Fe}{I} 630.2\,nm line recorded on 2011-Aug-06. The field-of-view is about $41\arcsec \times 41\arcsec$ cut from the original size of $57\arcsec \times 57\arcsec$ to fit with the Hinode SP image. The scale is presented in megameters [Mm]. Yellow boxes mark the location of a detected event in each particular dataset: event ID 523 for the Hinode SP image and event ID 53961 for the CRISP image, respectively. Green, red and blue contours indicate the apparent transverse magnetic flux, the negative longitudinal magnetic flux and the positive longitudinal magnetic flux, respectively.
    }
	\label{fig:1}
\end{figure}

\section{Observations and methods}
\label{sec:obs}

\subsection{Calibrated database}
\label{subsec:db}

To investigate the emergence of magnetic flux sheets associated with a growing granule seen in the continuum, we utilized data from two observatories: the Solar Optical Telescope \citep[SOT,][]{Tsuneta2008} of the Hinode satellite \citep{Kosugi2007} and the ground-based Swedish 1-m Solar Telescope \citep[SST,][]{Scharmer2003}. Spectropolarimetric observations were acquired by the Hinode Spectropolarimeter \citep[SP,][]{Hinodesp} and the Imaging Spectropolarimeter \citep[CRISP,][]{Scharmer2008}, respectively. All datasets were recorded during very low solar activity, near solar minima. Tables \ref{table:1} and \ref{table:2} list the specific datasets analyzed in this study. Figure \ref{fig:1} presents two representative snapshots, highlighting two distinct cases of emerging magnetic flux sheets.

\subsubsection{Hinode data}
\label{subsec:h_db}

To study the magnetic flux-sheet emergence within a growing granule seen in the continuum, we favour high-spatial-resolution (no binning) datasets of SP, having a pixel size of 0\farcs16. These datasets have a spectral coverage of 2.39\,\AA\,centred in the \ion{Fe}{I} 630.15 and 630.25\,nm lines (Land\'{e}-factor of $g_{\rm eff}=1.66$ and $g_{\rm eff}=2.5$, respectively) and a spectral sampling of 21.55\,m\AA\, in its normal operation mode. We select the datasets with less than or equal to a minute cadence, to catch the events which sometimes last less than 5 minutes and to identify the different stages of the event. A disadvantage of the SP datasets is that one obtains a relatively small field-of-view (FOV). In addition, one needs to exclude events which only take place partially within the FOV, therefore the effective FOV is further reduced. We try to alleviate the situation by collecting several datasets taken at different times and days, thereby, expanding the viewable area by combining multiple datasets. 

We obtained suitable data from one of the online repositories\footnote{\url{http://sdc.uio.no/sdc/}} choosing datasets close to the disc centre. In total, we have obtained over 15 hours of a FOV of approximately $2\farcs7  \times 41 \arcsec$ and about 7 hours of a FOV of approximately $4\farcs5  \times 41\arcsec$. The details of the chosen datasets can be found in Table \ref{table:1}. 
 
All datasets were calibrated using IDL routines provided by the instrument scientists and publicly available through the SolarSoft distribution. The SP calibration provides us with the apparent transverse and longitudinal magnetic flux density as calculated by \texttt{sp$\_$prep.pro} routine \citep[see][]{Lites2008}, which we mostly use in this work. This magnetic flux corresponds to the magnetic field as obtained for the area of the resolution element with filling factor one as if the magnetic field covered the entire resolution element.

\subsubsection{SST data}
\label{subsec:sst_db}

The SST data were obtained from two different observation runs in 2011 and 2019.

We use science-ready CRISP data from August 2011. The dataset consists of spectropolarimetric data taken in the \ion{Fe}{I} 630.15 and 630.25 lines sampled with 44\,m\AA\,steps at 30 wavelength points. The dataset additionally includes spectropolarimetric data at the \ion{Fe}{I} 557.6\,nm line, which is not used in this work. CRISP recorded spectral images for 47 minutes with a cadence of 28 seconds. The spatial sampling was at $0\farcs059 \times 0\farcs059$ per pixel with a FOV of about $53\arcsec \times 53\arcsec$. The dataset has been calibrated (dark-, flat-, wavelength-, and polarization calibration) and reconstructed with the Multi-Object Multi-Frame Blind Deconvolution \citep[MOMFBD,][]{Lofdahl2002,VanNoort2005} algorithm. The output includes Stokes-$I$, $Q$, $U$, and $V$ images at the 30 wavelength points. For more information about the dataset see also \cite{Stangalini2015,Stangalini2017,Ledvina2022}. We have produced total linear polarization and circular polarization maps taking into account both lines of the \ion{Fe}{I}\,630\,nm line pair. We use the resulting time series of polarization and continuum maps for further analysis. 

In addition, we use a dataset collected during a SOLARNET campaign executed in April 2019, which includes spectropolarimetric observations in the \ion{Fe}{I} 617.3\,nm line (Land\'{e}-factor of $g_{\rm eff}=2.5$) sampled with 15 spectral points symmetrically distributed around the line core ($\lambda_{0}$) in 35\,m\AA\,steps. This dataset also includes spectropolarimetric information from \ion{Ca}{II} 854.2\,nm and spectroscopic information from H$\alpha$ line, which are not used in this work. The FOV covered by the CRISP images was approximately $50\arcsec \times 50\arcsec$ sampled at a pixel size of $0\farcs059\times 0\farcs059$. The duration of the time sequence was $\sim$ 2 hours with a cadence 28 seconds. The data from CRISP have been calibrated (dark-, flat-, wavelength-, and polarization calibration) using the SSTRED package ~\citep{Lofdahl2018} as well as reconstructed with the MOMFBD algorithm. The science-ready data was then made available by the Institute for Solar Physics of Stockholm University. For more information about the dataset see also \cite{Fischer2020, DiazCastillo2024}. We produced total linear and circular polarization maps and aligned the data sets to each other taking the \ion{Fe}{I}\,617.3\,nm data as a reference. In addition to the polarization maps, we calculate the apparent longitudinal magnetic flux density from the Stokes V profiles using the weak field approximation with the same method applied in \cite{MartinezGonzalez2009}. 

\subsection{Magnetic flux sheet detection and characterization methods}
\label{subsec:method}

\begin{figure*} [h!]
	\centering
	\includegraphics[width=\textwidth]{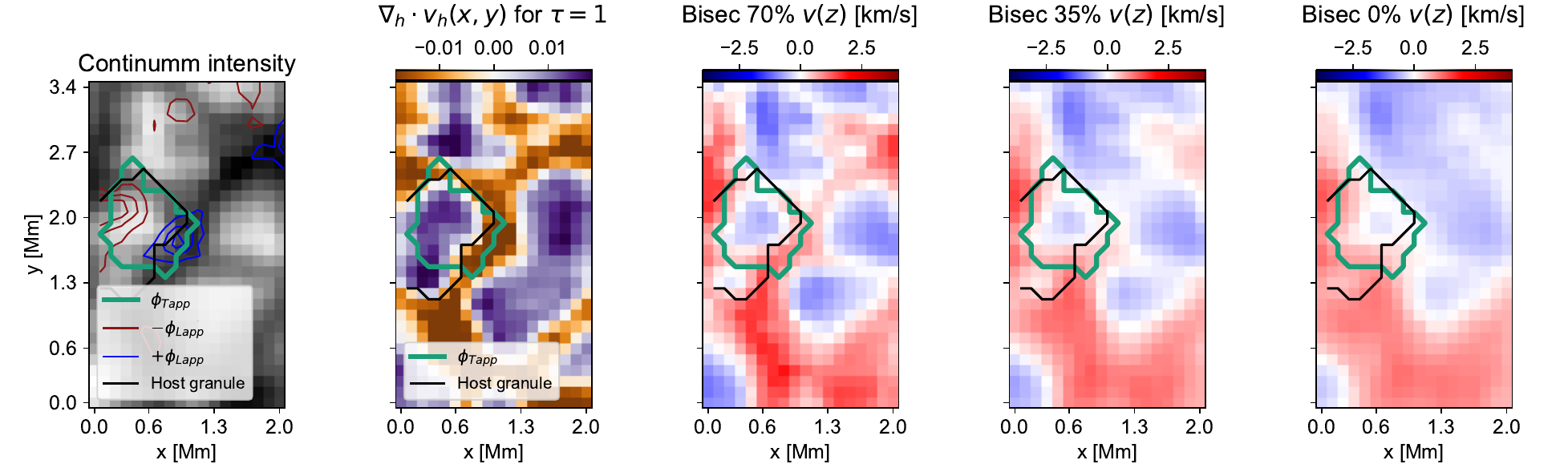}
	\caption{Magnetic flux sheet event with ID 352 recorded by Hinode SP on 2007-Sep-26 08:15UT. From left to right, the first panel shows the continuum intensity in the background with contours. Pine green contours mark a threshold of 130 Mx cm$^{-2}$ for the apparent transverse magnetic flux density patch $\phi_{\text{Tapp}}$. The blue (red) contours mark the positive $+\phi_{\text{Lapp}}$ (negative $-\phi_{\text{Lapp}}$) polarity of the longitudinal magnetic flux density with thresholds of 20 Mx cm$^{-2}$, 40 Mx cm$^{-2}$ and 60 Mx cm$^{-2}$. The black contours mark a single closed contour of zero-divergence defined as the boundary of the host granule. The second panel shows the divergence of the apparent horizontal velocity field marking the corresponding location of the sheet represented by the transverse magnetic flux density patch $\phi_{\text{Tapp}}$ (pine green contour) and the boundary of the host granule (black contour). Finally, the third, fourth and fifth panels show the bisector velocity field for the three different depth positions: 70\%, 35\% and core (0\%) of the \ion{Fe}{I} 630.15\,nm line corresponding roughly to the low, middle and high photosphere, respectively.}
	\label{fig:2}
\end{figure*}

The main criteria used for identifying events of magnetic flux sheet emergence is the evidence of the simultaneous evolution of a transverse magnetic flux patch and a host granule. Therefore, the first task is to find patches of transverse magnetic flux. For the detection and tracking of the transverse magnetic flux, we use the Yet Another Feature Tracking Algorithm (YAFTA) code \citep{2003WandL}, which is applied to the apparent transverse magnetic flux densities calculated using both lines of the Hinode SP data and the total linear polarization maps of the \ion{Fe}{I} 630.25 and 617.3\,nm of the SST CRISP datasets of 2011 and 2019, respectively. The minimum size of around 0.1 Mm$^2$ for the transverse magnetic flux patches and threshold parameters are set separately for the Hinode SP or SST CRISP datasets. This step can then be performed fully automatically without visual guidance and allows us to obtain a group of potential events. For the Hinode SP, we set a threshold of 130 Mx cm$^{-2}$ for the apparent transverse magnetic flux density corresponding to three times the noise level ($\sim$\,40 Mx cm$^{-2}$). For the SST CRISP dataset, we set a threshold of $5\times10^{-3} I_c$ of the total linear polarization corresponding to three times the noise level ($\sim1.7\times10^{-3} I_c$). 

The second task is to link the evolution of the transverse magnetic flux to the evolution of a host granule co-located with the transverse magnetic flux patch and evolving simultaneously with it. We explore the granular pattern associated with each event by comparing the temporal evolution of the patches of transverse magnetic flux with spatially aligned continuum images. During our analysis, we found that the automated assignment of a single granule based solely on the variation of the intensity was in some cases rather difficult due to the complexity and fine structure of the convection pattern. Hence, to effectively associate the evolution of a unique granule in a semi-automatic procedure, we implement a method that combines the horizontal velocity field and its divergence, and the line-of-sight (LOS) or vertical velocity field. 

To estimate the horizontal velocity distribution of the plasma flow, we use the DeepVel neural network model \citep{AsensioRamos2017} applied to continuum images. DeepVel is a deep learning model based on a fully convolutional neural network architecture that predicts instantaneous horizontal velocities at different heights in the lower solar atmosphere. It directly estimates these velocities pixel-by-pixel from pairs of consecutive continuum images, without time or space averaging. This distinguishes DeepVel from other methods that require such averaging for accurate results such as the standard local correlation tracking methods \citep{November1988,Fisher2008}. Synthetic data generated by suitable magnetohydrodynamics (MHD) models of the solar atmosphere are used to train DeepVel. These models provide images of the solar surface and the corresponding true velocities at three different heights. DeepVel learns to associate these images with the underlying velocities through a deep learning process. The accuracy of DeepVel's predictions is closely tied to the specific characteristics of the training data, thus to estimate velocities at different spatial resolutions, observing cadences, or locations on the Sun, the DeepVel model should be retrained using synthetic data that reflects these specific conditions \citep{Tremblay2018}.

For the Hinode SP dataset, we perform a training and testing procedure using two different CO$^5$BOLD simulations described in \cite{Vigeesh2017}, one hydrodynamic (non-magnetic) and the other magnetohydrodynamic simulation (initial uniform vertical magnetic field of 50\,G and advanced over a magnetic field redistribution timescale). The computational domain has a horizontal size of around $38 \times 38$\,Mm with a cell size of 80\,km and discretized on $480 \times 480$ horizontal grid cells. We use snapshots of intensity at the different cadences of the Hinode SP datasets, which have been degraded with the corresponding PSF profile. We use the bolometric intensity at $\tau = 1$ of the simulations, which reproduces the behaviour of the observed continuum, and match the histogram distribution of the simulated intensity with the observed continuum intensity. For training and validation of the model, we randomly extract 64000 pairs of consecutive patches of $40 \times 40$ pixels from the PSF-degraded continuum snapshots together with the corresponding horizontal velocity field at three different $\tau$ values: $\tau = 1, 0.1, 0.01$ from the simulation. As in \cite{AsensioRamos2017}, all inputs are normalized to the median intensity of the quiet Sun and velocities to the interval [0, 1] using training set bounds. The model architecture was implemented using PyTorch modules and the training is carried out by optimizing the squared difference between the output of the network and the velocities in the training set with the Adam stochastic first-order gradient-based optimization algorithm \citep{Kingma2014} using a learning rate of 0.001 and batches of 64 samples during 50 epochs.

For the SST dataset, we use the default and pre-trained model from \cite{AsensioRamos2017}. This model was trained on synthetic magneto-convection simulations and tested on \textsc{Sunrise}~I/IMAX \citep{Barthol2011,Martines2011}; it is applicable to CRISP data successfully as there is only a slight difference between the \textsc{Sunrise}~I/IMAX data pixel size and the CRISP pixel size. We note, however, that the model was trained considering snapshots with a cadence of 30 seconds, which may lead to an underestimation of velocities compared to the 28-second cadence of SST CRISP data used in this work. 

While the horizontal velocities inferred by the DeepVel neural network are model-dependent, it accurately reconstructs the flow orientation and amplitude at subgranular and granular scales when using a training and validation input generated from different numerical simulations \citep{Tremblay2018}.

To estimate vertical velocity fields, we calculate the bisector velocity as a proxy for LOS velocities at specific line depths thus sampling different heights in the photosphere. Due to the different spectral resolutions of the Hinode SP and CRISP and the different formation heights of the lines, we select lines and define three line depths that roughly characterize the low, middle and high photosphere. For the Hinode SP spectra, we select depth positions at 70\%, 35\% and core (0\%) of the \ion{Fe}{I} 630.15\,nm line for the low, middle and high photosphere, respectively. For SST CRISP, we select depth positions at 60\%, 30\% and core (0\%) of the \ion{Fe}{I} 630.15\,nm line (2011 dataset) and of the \ion{Fe}{I} 617.3\,nm line (2019 dataset) for the low, middle and high photosphere, respectively. We also perform p-mode correction to the full sequence of the datasets.

Based on the description of \cite{MorenoInsertis2018}, the emergence of magnetic flux sheets can take place in the early stages of the host granule, which are characterised by rapid and expanding plasma flows. We therefore use the positive transverse divergence of the flow to determine the area of the host granule. In addition to the horizontal plasma flow, the vertical plasma flow is expected to be upward-directed inside expanding granules and gradually weaken at higher altitudes.

Considering the above, we define two initial criteria which characterize a granule hosting a magnetic flux sheet: 1) closed regions of positive divergence and 2) closed regions of upward velocities, both co-located with the patch of transverse magnetic flux. To apply such criteria, we apply the DeepVel models to the time series of continuum intensity images of all datasets, to retrieve the apparent horizontal velocity field ${\bf v}_h$ and then calculate its divergence, $\nabla_h \cdot {\bf v}_h(x,y)$. We select the plasma flows at the lowest height in the photosphere to identify the host granule, that is, the apparent horizontal velocity field at $\tau = 1$ provided by DeepVel and the vertical velocity field at the 70\% and 60\% of the line depth provided by the bisector analysis for Hinode SP and SST CRISP, respectively. 

Our method applies the first criteria automatically as follows: 1) locate the superposition of regions of positive divergence of the apparent horizontal velocity field with the patch of transverse magnetic flux and 2) define the host granule boundaries using a single closed contour of zero-divergence. We note that some events show a complex granulation pattern in their surroundings and that spatial resolution makes the zero-divergence criteria too sensitive to contrast changes resulting in merged granules. Thus, we manually inspect the distribution of the vertical velocity field inside the mask to correct merging granules. In some cases, we apply both criteria strictly even though this may result in a relatively smaller granule. In addition, many of the events were associated with exploding granules and granular lanes (see Sect.\,\ref{sec:res} for details on these phenomena and findings). This leads to cases in which a part of the granule (a newly split or sectioned area) starts to expand again and develops its own convective pattern. In some cases, the expanding section is viewed as the host granule when is responsible for surfacing the associated magnetic flux.

Finally, the third task is to determine if the newly arriving transverse magnetic flux patch is flanked by appearing and diverging opposite polarity patches of longitudinal magnetic flux density, making this a true emerging event. We defined the longitudinal magnetic flux patches associated with the footpoints by proximity to the sheet. We select the longitudinal magnetic flux patches within a radial distance corresponding to a circle centred on the sheet, with a maximum area of two times the maximum area of the sheet, and we define the boundaries of the longitudinal magnetic flux patches with a threshold. To establish the threshold values, we perform a similar calculation of the correspondent magnetic flux density retrieved by a particular photon noise as presented by \cite{MartinezGonzalez2009}. We set 10 Mx cm$^{-2}$ and 15 Mx cm$^{-2}$ for Hinode SP and SST CRISP datasets, respectively, corresponding to three times the photon noise in each case. 

In both the Hinode SP and SST CRISP datasets, we visually inspected the results and, when necessary, manually re-adjusted boundaries. This visual inspection was crucial for confirming the effective presence of magnetic flux sheet cases. Feature tracking methods like YAFTA identified and tracked all transversal magnetic flux patches in the considered datasets, encompassing all potential cases of loop-like magnetic flux emergence. Hundreds of such patches were detected. However, to accurately identify sheet emergence, in addition to the established criteria, we visually verified all cases to minimize the occurrence of false detections.

Figure\,\ref{fig:2} shows an example of a magnetic flux sheet emergence event in the Hinode SP dataset characterized by our method. All panels correspond to a single time step where the magnetic flux sheet has maximum extent over its host granule.    

\section{Results} 
\label{sec:res}
\subsection{Occurrence rate, description and properties of events} 
\label{main_prop}

\begin{figure*}
    \centering
    \begin{subfigure}{0.3\textwidth}
        \includegraphics[width=\linewidth]{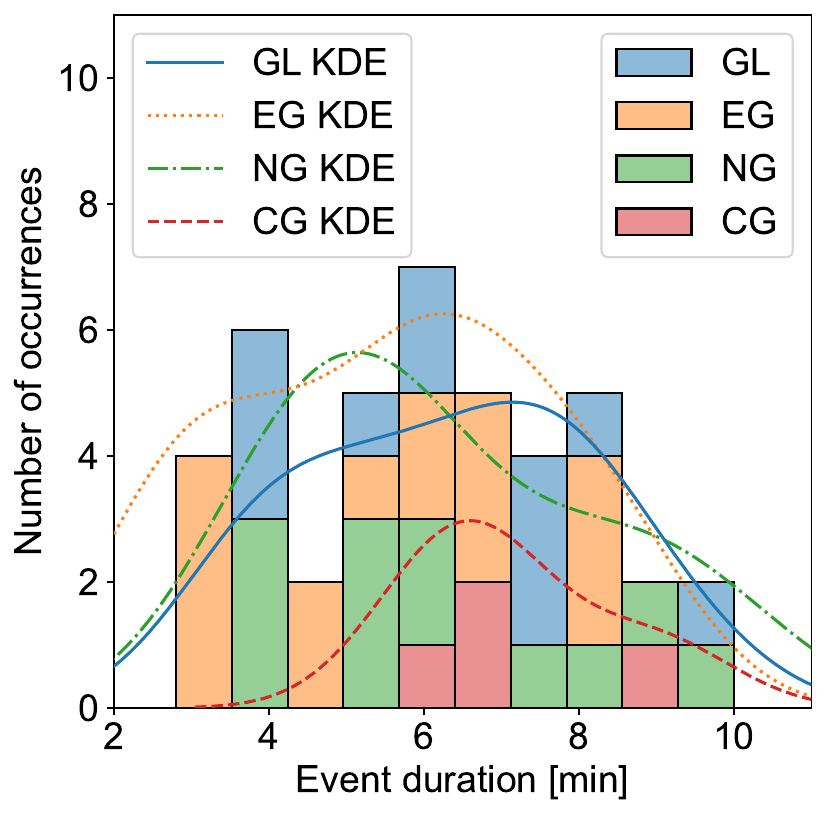}
        \caption{Event duration}
        \label{fig:3.1}
    \end{subfigure}
    \hfill
    \begin{subfigure}{0.3\textwidth}
        \includegraphics[width=\linewidth]{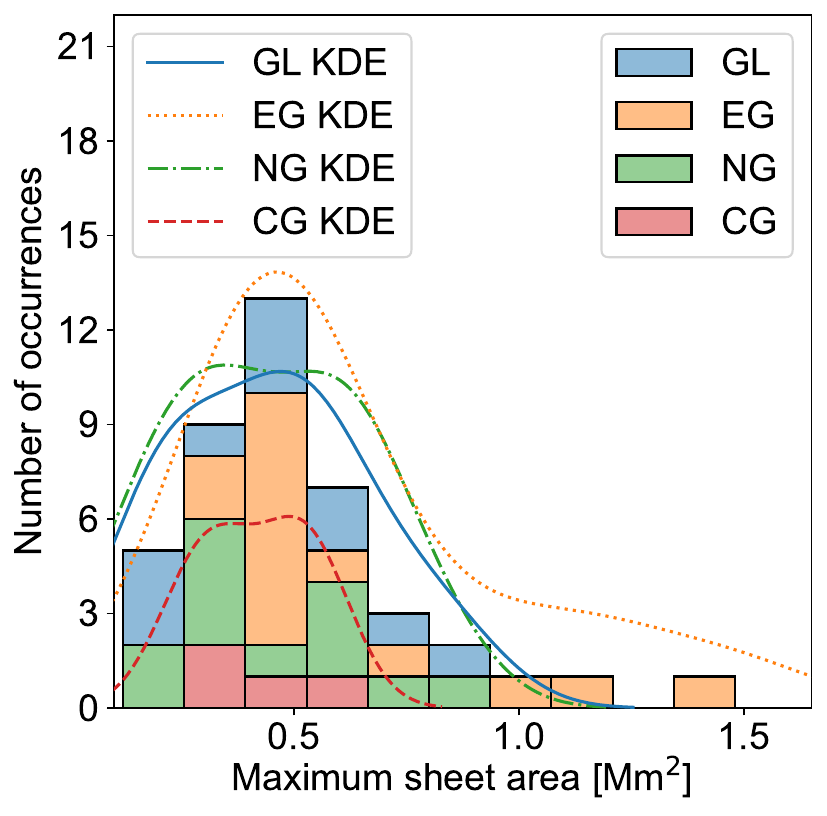}
        \caption{Magnetic flux sheet maximum area}
        \label{fig:3.2}
    \end{subfigure}
    \hfill
    \begin{subfigure}{0.3\textwidth}
        \includegraphics[width=\linewidth]{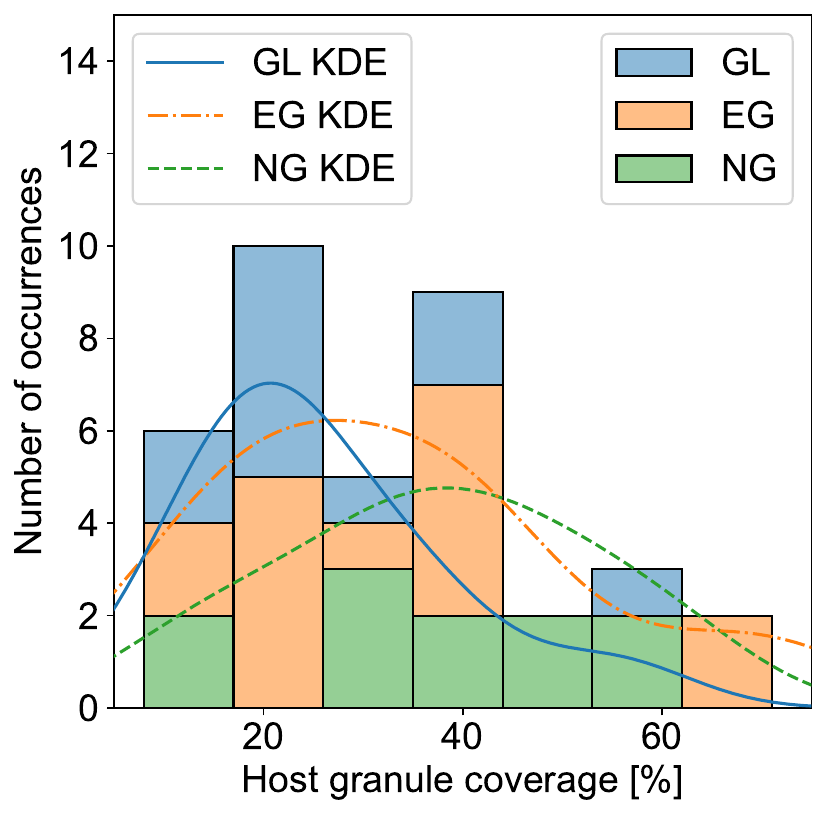}
        \caption{Host granule coverage}
        \label{fig:3.3}
    \end{subfigure}
    \caption{Histograms of the duration of each event (left panel), the apparent maximum area reached by the magnetic flux sheet (central panel) and the proportion of a host granule's area covered by its associated magnetic flux sheet (right panel). The granulation type is coded by different colours, as follows: exploding granules (EG): yellow bars and lines; granular lanes (GL): blue bars and lines; nascent granules (NG): green bars and lines; and complex granulation (CG): red bars and lines. The complex granulation is only included in panels (a) and (b). All histograms include the kernel density estimate (KDE) providing complementary information about the possible shape of the distribution. The actual values are included in Cols. 4-5 of Table \ref{table:3} and Col. 4 of Table \ref{table:4}}
    \label{fig:3}
\end{figure*}

After detecting events as described in Sect. \ref{subsec:method} and discarding questionable candidates (mostly those located at the edges of the FOV and those with incomplete evolution information), we confirm 42 events as positive identifications (see full databases in Tables \ref{table:3} and \ref{table:4}). This number likely represents a lower limit of the actual occurrence rate of magnetic flux sheets. The numerical simulations by \cite{MorenoInsertis2018} estimate $0.3$ to $1$ events per day and Mm$^2$. However, as the authors also note, this estimate is based on a small sample found in the simulations and might not be representative. Assuming the lower estimate of 0.3 events per day and Mm$^2$, we would expect to detect approximately 45 events in our database, which is a number comparable to the 42 events we identified. 

As an initial classification that helps to characterise the magnetic sheet events, we define four types of granular phenomena linked to the host granule, where the emergence of the magnetic flux sheet occurs: exploding granules (EG), nascent granules (NG), granules with granular lanes (GL), and complex granulation (CG). We found that the first three types of granular phenomena associated with magnetic flux sheet appearance are roughly equally represented in our sample.

The most frequent are magnetic flux sheet events associated with exploding granules, with 15 cases identified. Exploding granules appear in continuum images as rapidly expanding, often large granules that develop a dark core or dot \citep{Hirzberger2001,Roudier2001,Ellwarth2021}. This dark core indicates a developing downflow within the granule, which eventually evolves into a new intergranular lane. In addition to their morphological features, exploding granules have been associated with nearby magnetic flux dynamics \citep{DePontieu2002,Zhang2009}. In particular, \cite{Palacios2012} and \cite{Guglielmino2020} studied the emergence of extended magnetic structures in exploding granules observed by \textsc{Sunrise}~I/IMAX, carrying total magnetic fluxes of the order of $10^{18}$ Mx. Using numerical simulations, ~\cite{Rempel2018} reported on the process of amplification of magnetic flux by turbulent shear in the newly forming downflow within the exploding granule. This process results in the formation of extended magnetic structures nearby, with lengths comparable to the granule scale. 

The next phenomena in the ranking are nascent granule events, with 12 cases, and granules with granular lanes, with 11 cases. In the nascent granule events, the magnetic flux sheet appears nearly simultaneously with the host granule itself and evolves with it during its fast expansion phase. These are the cases mentioned in the numerical paper of \citet{MorenoInsertis2018} as a probable cause of the formation of magnetic flux sheets. The average size of nascent granules in this sample is approximately 0.8 Mm$^2$, which is smaller than the average size of host granules classified as exploding granules and host granules with granular lanes (1.7 Mm$^2$). However, the average maximum area reached by nascent granules in the sample is around 1 Mm$^2$, a characteristic size of standard granules. This suggests that these sheets tend to disappear (or become undetectable) in the advanced phases of the granular evolution.

Granular lanes, on the other hand, appear in continuum images as dark lanes moving within a granule, preceded by a bright rim. They have an arch-like shape, with their endpoints still connected to the intergranular lane as they move apart. Numerical simulations by \cite{Steiner2010} suggest that these features are the visible signatures of horizontal vortex tubes beneath the solar surface, which can be associated with magnetic fields in the photosphere, as confirmed observationally by \cite{Fischer2020}. 

We identified four cases with complex granulation patterns (CG) where the host granule could not be clearly determined. As a result, we excluded these cases from the statistics on the host granule coverage and the plasma dynamics of host granules. 

Figure \ref{fig:3} summarizes the temporal and spatial properties of the events concerning the associated granular type. Columns 4-5 of Table \ref{table:3} and Col. 4 of Table \ref{table:4} show the absolute value for each quantity analysed. We found no clear relationship between the duration of magnetic flux sheet events and the associated granular type, as shown in Fig.\,\ref{fig:3.1}. The duration of the events ranges from 3 to 10 minutes, with an average of 6 minutes. Similarly, the maximum area reached by the transversal magnetic flux density patch, from now on termed magnetic flux sheet area, does not show a clear correlation with the granular type, as shown in Fig.\,\ref{fig:3.2}, with most magnetic flux sheets exhibiting areas between 0.25 and 0.75 Mm$^2$. 

However, the proportion of a host granule's area covered by its associated magnetic flux sheet (host granule coverage) appears to be linked to the granular type (see Fig.\,\ref{fig:3.3}). Nascent and smaller granules (< 1 Mm$^2$) are generally covered to a larger extent by the magnetic flux sheet compared to larger granules (> 1 Mm$^2$) (green histogram in Fig.\,\ref{fig:3.3}). Magnetic flux sheets associated with granular lanes typically cover a smaller area in the host granule (areas between 1 Mm$^2$ to 2 Mm$^2$), as they tend to be elongated and located near the edges of granules, which prevents them from covering it completely (blue histogram in Fig.\,\ref{fig:3.3}). The distribution of the host granule coverage for events associated with exploding granules seems broader (yellow histogram in Fig.\,\ref{fig:3.3}). This could be due to two factors: (1) a wider range of the sheet sizes (blue histogram in Fig.\,\ref{fig:3.2}) or a wide range of host granule sizes (areas from 1 Mm$^2$ to 5 Mm$^2$), which will depend on whether the host granule is the entire exploding granule or a fragment of it when the sheet reaches its maximum. We found that the maximum coverage measured in our event list reached only 70\%, which could be related to several factors: the criteria used to define the boundaries of the host granule or the intrinsic properties of the granule or sheet. 

In the following subsections, we examine the evolution of selected cases for each granulation phenomenon. The simplest, standard "textbook-like" cases of flux sheet emergence in our sample are found for the category of nascent granules (Sect.\,\ref{SG} and, in particular, Fig.\,\ref{fig:7}). Yet, we start describing individual flux sheet emergence events with examples of exploding granules since they are the most frequent category in the sample. We also illustrate the diverse characteristics of emerging magnetic flux sheets and highlight, from the beginning, various difficulties in their identification.

\begin{figure*}
	\centering
	\includegraphics[width=\textwidth]{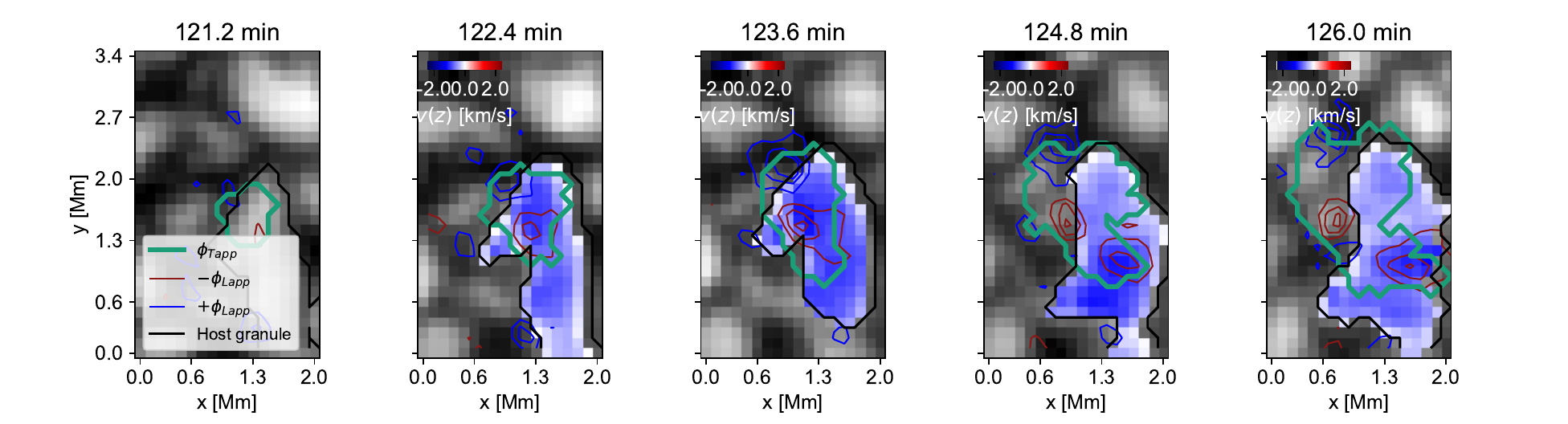}
	\caption{Magnetic flux sheet emergence event alongside the expansion of its host granule, which fragments as an exploding granule. The image series shows the continuum maps reconstructed from Hinode SP slit data. The times are indicated above the images. Contours mark the apparent transverse (pine green) and longitudinal (red contours for the negative polarity and blue contours for the positive polarity) magnetic flux density associated with the event. The magnetic flux density thresholds are defined as in Fig.\,\ref{fig:2}. Black contours mark the host granule border. Colour maps within the host granule show the distribution of the bisector velocity at 70\% of the line depth.}
	\label{fig:4}
\end{figure*}

\begin{figure*}
	\centering
	\includegraphics[width=\textwidth]{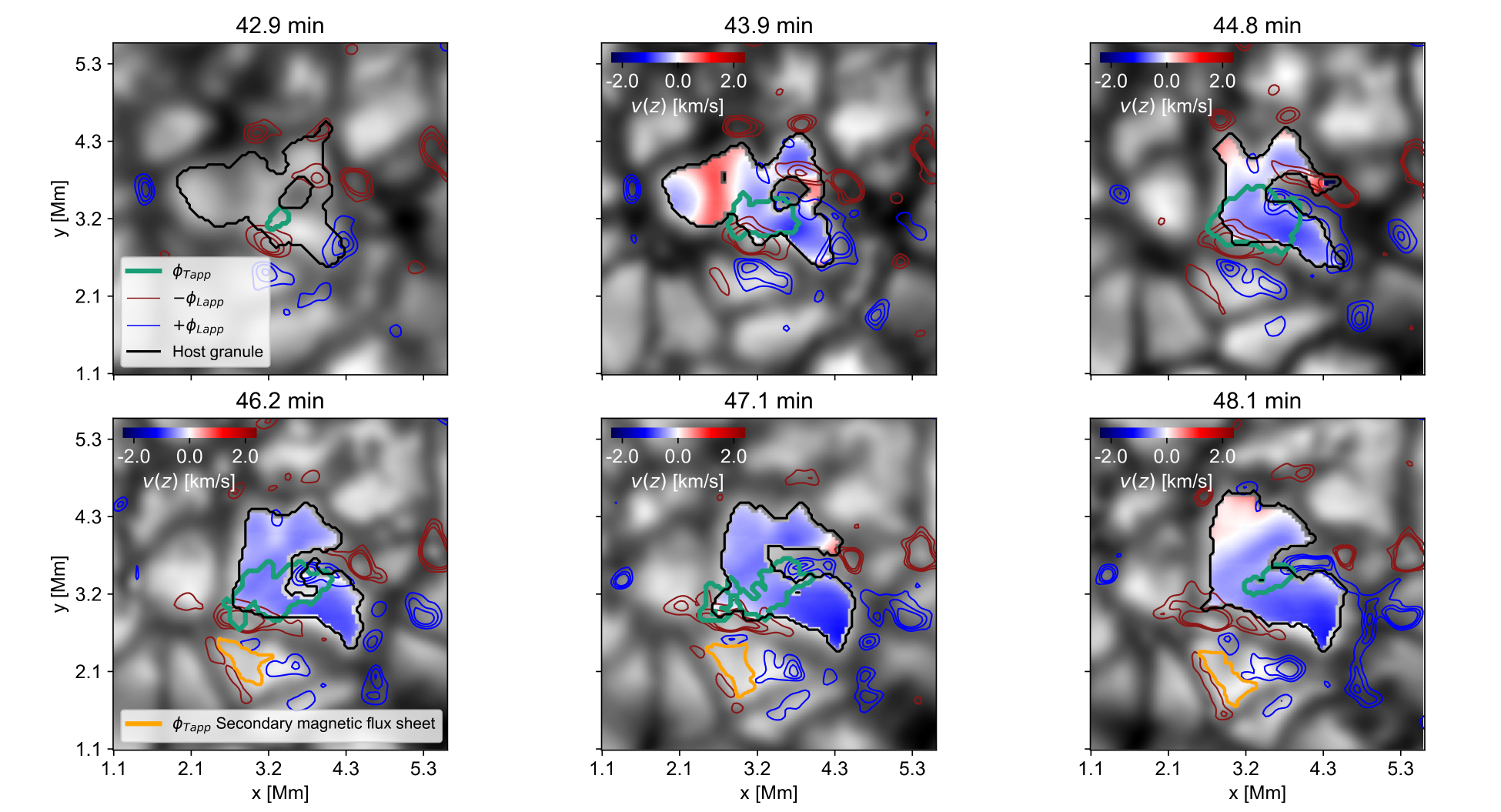}
	\caption{A second case of magnetic flux sheet emergence in an exploding granule. The image series shows the continuum close to the \ion{Fe}{I} 617.3\,nm line recorded by CRISP. Plot setup is defined as in Fig\,\ref{fig:4}. The blue (red) contours mark the positive (negative) polarity of the longitudinal magnetic flux density with thresholds of 40 Mx cm$^{-2}$, 60 Mx cm$^{-2}$ and 80 Mx cm$^{-2}$. Black contours mark the host granule border. A secondary magnetic flux sheet, depicted by orange contours, was detected near and concurrently with the event within the exploding granule. Colour maps within the host granule show the distribution of the bisector velocity at 60\% of the line depth.}
	\label{fig:5}
\end{figure*}

\subsubsection{Exploding granule events}
\label{EG}

Figure \ref{fig:4} shows the emergence of a magnetic flux sheet associated with an exploding granule. This event, labelled ID 801, was detected by Hinode SP during the 2007-Sep-25 12:59UT run. The magnetic flux sheet (pine green contour) becomes visible at the boundary of the exploding granule (black contour) at $t = 121.2\,\text{min}$. The characteristic dark dot is observed around $(x,y) = [1.0,0.9]$ at $t = 122.4\,\text{min}$ and $t = 123.6\,\text{min}$ while the magnetic flux sheet expands. Along with the sheet, both footpoints of longitudinal magnetic flux are visible as small patches flanking the sheet (red contours for the negative polarity and blue contours for the positive polarity). The main fragment of the exploding granule is identified as the host granule (black contour), characterized by upflows of $-2\,\text{km s}^{-1}$ (blue pixels on the colour map within the host granule). Over the five-minute evolution, the magnetic flux sheet quadruples in size, the footpoints expand and intensify, and the host granule appears to grow in area. In this event, the positive-polarity footpoint remains anchored to the intergranular region from the start of the emergence, while the negative-polarity footpoint appears within the host granule. After $t = 124.8\,\text{min}$, the negative-polarity footpoint fragments into two patches and a second patch of positive polarity appears surrounding the left flank of the magnetic flux sheet patch. This observation is similar to the organization of the longitudinal magnetic flux described by \citet{MorenoInsertis2018} during the later stages of magnetic flux sheet emergence. In their study, the two circular polarization patches, representing the footpoints, gradually move apart and stretch over time, eventually reaching the intergranular lanes.

Figure \ref{fig:5} depicts the complete evolution of a magnetic flux sheet event within an exploding granule observed by SST CRISP. This event, identified as ID 91772, is detected in the 2011 dataset and is associated with a "double-event" apparently related to a large-scale magnetic structure accumulating beneath the photosphere and emerging in sections. A small transverse magnetic flux patch appears below the typical central dark dot in the exploding granule at $t = 42.9 \text{min}$ (pine green contour). Around the main transverse magnetic flux patch, non-negligible signals of linear polarization located within neighbouring granules and associated with the distribution of the longitudinal magnetic flux are detected. At $t = 43.9\,\text{min}$, the granule begins to disintegrate, forming intergranular lanes associated with downflows of $1\,\text{km s}^{-1}$ (red pixels on the bisector velocity map within the host granule). At the same time, the transverse magnetic flux patch remains in a region dominated by upflows (blue pixels on the bisector velocity map within the host granule). One intergranular lane appears on the granule's left side far from the central dot, while the other emerges on the right connected to the central dot. By $t = 44.8\,\text{min}$, the transverse magnetic flux patch reaches its maximum extent and a fragment of the original granule becomes the host granule (black contour). The longitudinal magnetic flux patches linked to the footpoints of the sheet are located within the intergranular lane on the host granule surroundings, where they undergo splitting and merging with nearby longitudinal magnetic flux concentrations over the next three minutes. From $t = 46.2\,\text{min}$ onwards, a secondary event, identified as ID 94010, appears within a neighbouring granule around $(x,y) = [3.1,2.1]$ (contour orange in Figure \ref{fig:5}). The surrounding longitudinal magnetic flux in the FOV shows a complex distribution associated with this event, evidencing a potential fragmentation of a large-scale magnetic structure while emerging from below the solar surface.

\begin{figure*}
	\centering
	\includegraphics[width=0.92\textwidth]{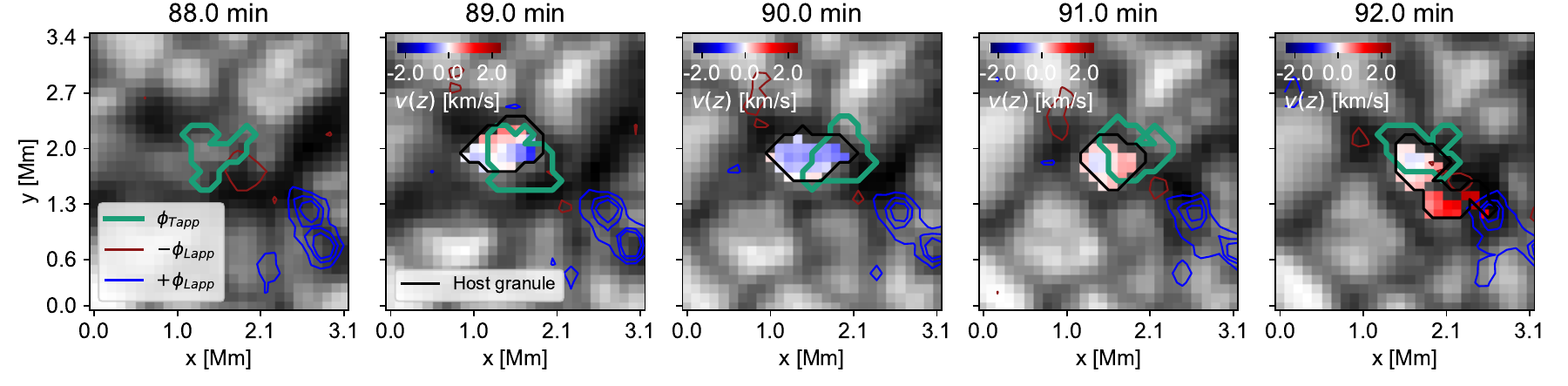}
	\caption{Case of a magnetic flux sheet emergence in a nascent and relatively small granule. The image series shows the continuum maps constructed from Hinode SP slit data. Plot setup and information are as shown in Fig\,\ref{fig:4}.}
	\label{fig:6}
\end{figure*}

\begin{figure*}
	\centering
	\includegraphics[width=\textwidth]{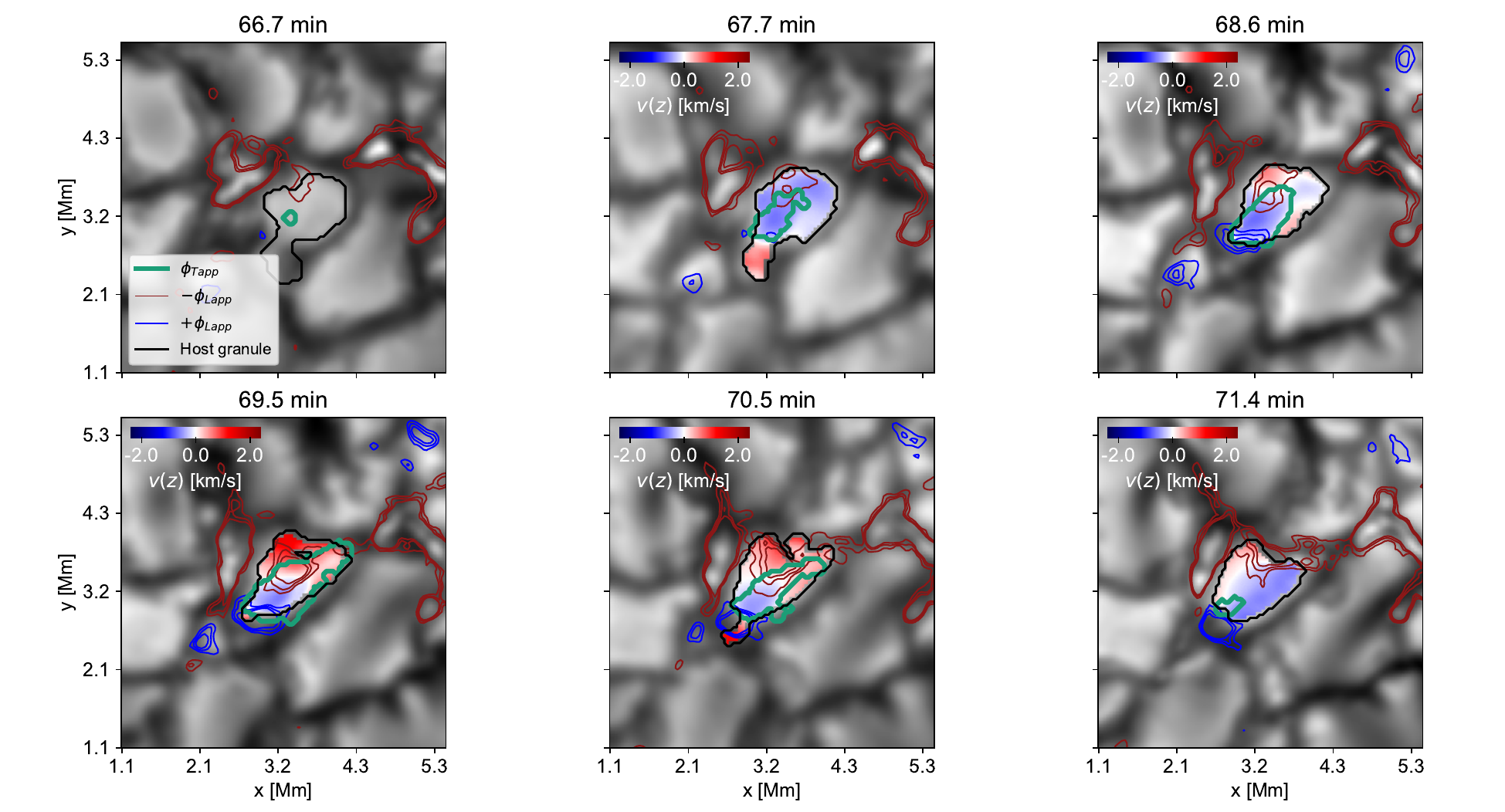}
	\caption{A second case of magnetic flux sheet emergence in a nascent granule. The image series shows the continuum close to the \ion{Fe}{I} 617.3\,nm line recorded by CRISP. Plot setup and information are as shown in Fig\,\ref{fig:5}.}
	\label{fig:7}
\end{figure*}

\subsubsection{Nascent granule events}
\label{SG}

The event depicted in Fig.\,\ref{fig:6} showcases a magnetic flux sheet that appears over a nascent, isolated and relatively small granule. This event, labelled ID 458, was detected by Hinode SP during the 2007-Oct-22 18:34UT run. We measure a maximum area of the transverse magnetic flux patch of 0.4 Mm$^2$ reached at $t = 89.0\,\text{min}$ (pine green contour), just a minute after the initial detection and persisting for the following three minutes. The area of the host granule (black contour) and the area of the magnetic flux sheet (pine green contour) are similar and do not change after $t = 89.0\,\text{min}$. The vertical velocity within the host granule at $t = 89.0\,\text{min}$ and $t = 90.0\,\text{min}$ is characterized by upflows between $-0.5\,\text{km s}^{-1}$ to $-1.25\,\text{km s}^{-1}$ (blue pixels on the colour map within the host granule). These upflows transitioned to descending flows during the final two minutes, ultimately leading to the disappearance of both the sheet and the host granule, giving rise to an intergranular region. This event represents one of the smallest cases detected, exhibiting a relatively weak magnetic flux contribution. The relatively low spatial resolution of the Hinode dataset, coupled with the faint appearance of the event, significantly complicates the identification of its footpoints. As a result, we only consider a single footpoint with negative polarity, detected just above the detection threshold at the onset of the event.

\begin{figure*}
	\centering
	\includegraphics[width=\textwidth]{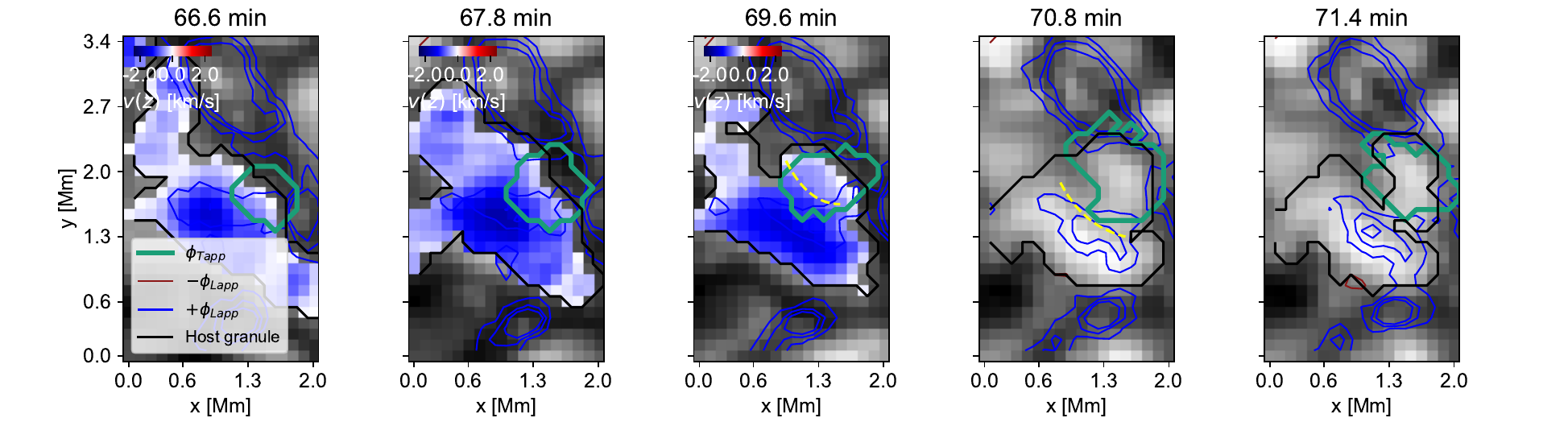}
	\caption{Case of a magnetic flux sheet emergence associated with a granular lane. The image series shows the continuum maps constructed from Hinode SP slit data. Plot setup and information are as shown in Fig\,\ref{fig:4}. The yellow dashed curves in $t = 69.6\,\text{min}$ and $t = 70.8\,\text{min}$ highlight the granular lane signature.}
	\label{fig:8}
\end{figure*}

\begin{figure*}
	\centering
	\includegraphics[width=\textwidth]{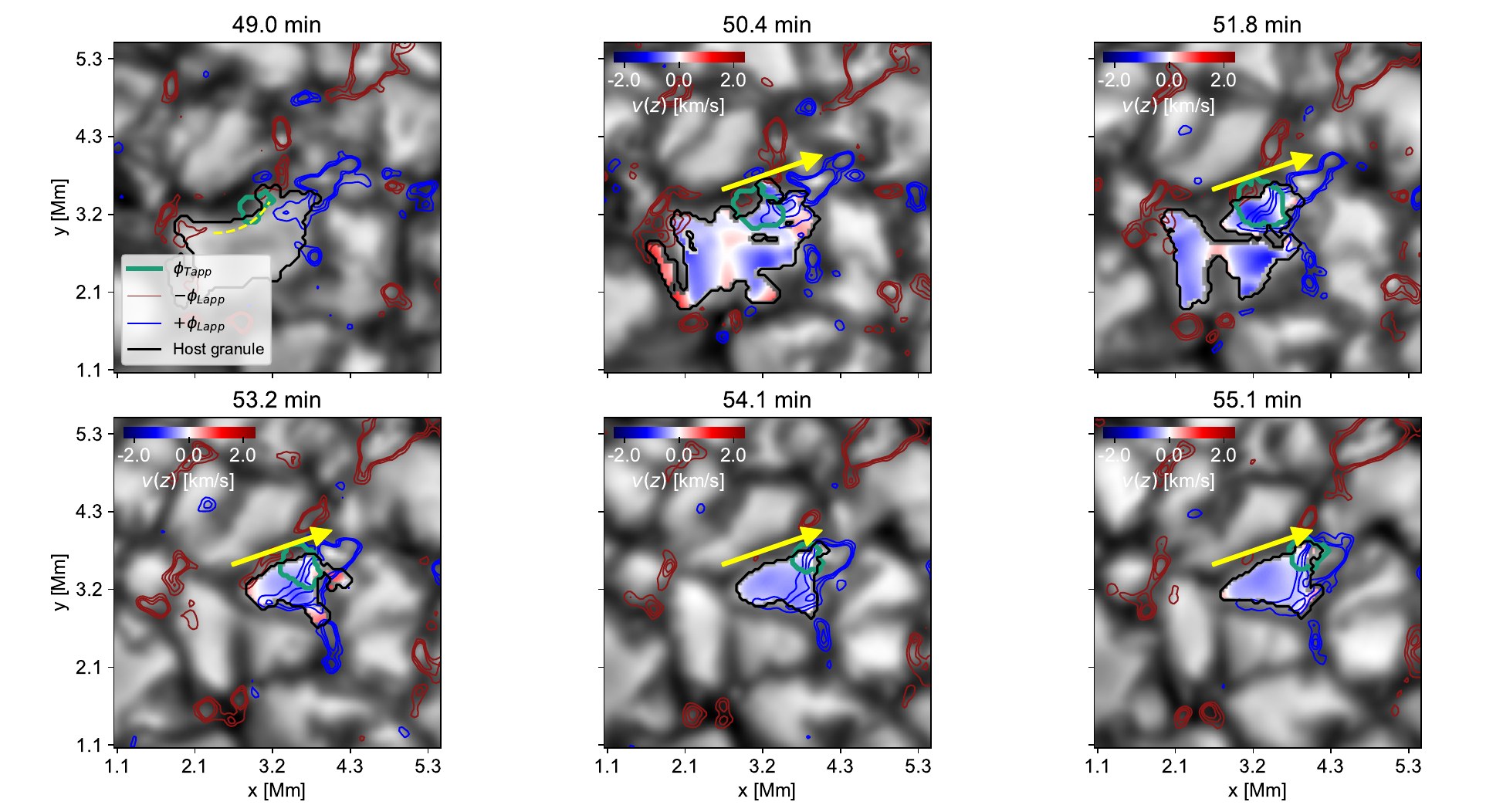}
	\caption{Case of a magnetic flux sheet emergence associated with a granule with granular lanes. The image series shows the continuum close to the \ion{Fe}{I} 617.3\,nm line recorded by CRISP. Plot setup and information are as shown in Fig\,\ref{fig:5}. The yellow dashed curve in $t = 49.0\,\text{min}$ highlights a granular lane signature. The yellow arrow marks the direction of displacement of the magnetic flux sheet.}
	\label{fig:9}
\end{figure*}

Figure \ref{fig:7} depicts the evolution of a magnetic flux sheet event within a uniform granule in its fast-expanding phase observed by SST/CRISP. This event, identified as ID 58352 and detected in the 2019 dataset, is a clear example of a magnetic sheet covering a considerable portion of its host granule. Although the host granule appears well-established in the initial continuum images, the vertical velocity field exhibits characteristics reminiscent of the early expansion phase of a nascent granule. The transverse magnetic flux patch appears within the host granule with an area below 0.1 Mm$^2$ at $t = 66.7\,\text{min}$, later on growing up to $0.5$ Mm$^2$ at $t = 67.7\,\text{min}$. The longitudinal magnetic flux patches corresponding to the footpoints appear in the flanks of the sheet, and the host granule develops an intergranular lane at the bottom of the sheet co-located with the positive-polarity footpoint (blue contours at $t = 67.7\,\text{min}$ and $t = 68.6\,\text{min}$). On the opposite side, the negative-polarity footpoint appears within the granule in a region of apparent downflow. At $t = 69.5\,\text{min}$, the magnetic flux sheet reaches its maximum extent of 0.68 Mm$^2$ and covers around 60\% of the host granule. One can observe that during that time, the host granule has a localized region of upflow fully covered by the sheet (blue pixels on the colour map in the bottom part of the host granule) and regions of downflow mostly localized at the edges on the upper part of the granule. In particular, the region with larger downflow velocities is co-located with the negative-polarity footpoint that moves towards the upper zone of the host granule until it reaches the intergranular region at $t = 71.4\,\text{min}$.

\subsubsection{Granular lane events}
\label{GL}

We observed that events associated with granular lanes often coincide with granule fragmentation or in the formation of a new convective pattern. Figures\,\ref{fig:8} and \,\ref{fig:9} exemplify these cases, the first one identified as ID 440 and detected by the Hinode SP during the 2007-Oct-06 08:01UT run and the second one identified as ID 39266 detected in the SST CRISP dataset of 2019, respectively. 

Figure \ref{fig:8} depicts the evolution of a magnetic flux sheet associated with a granular lane signature. The patch of transverse magnetic flux appears at the edge of an emerging granule at $t = 66.6\,\text{min}$. It expands over the next two minutes, reaching a size of 0.58 Mm$^2$ but only covering 20\% of the total area of the host granule. Signatures of the granular lane are visible at $t = 69.6\,\text{min}$ and $t = 70.8\,\text{min}$ (yellow dashed curves). The sheet is co-located with a dark area of relatively small upflow velocities (see the yellow dashed curve at $t = 69.6\,\text{min}$), preceded by a bright rim around $(x,y) = [1.7,2.0]$, that eventually lead to the formation of a new convective pattern after $t = 71.4\,\text{min}$. This event occurs near a strong network element, which does not seem to be associated with the sheet, therefore, we take as the effective footpoint the longitudinal magnetic flux patch of positive polarity closest to the sheet, which is only observed in the first two minutes of evolution in $(x,y) = [2.0,1.5]$. In this case, the footpoint of negative polarity is not observed. 

Figure \ref{fig:9} shows an example of a magnetic flux sheet with granular lane signatures evolving within a complex-shaped granule that eventually fragments. The patch of transverse magnetic flux appears at $t = 49.0\,\text{min}$ in an arch-like border of the host granule in its upper part similar to an extended granular lane, around $(x,y) = [3.0,3.2]$ (yellow dashed curve). At $t = 50.4\,\text{min}$, the host granule starts to fragment while the patch of transverse magnetic flux expands. The patch of transverse magnetic flux reaches its maximum extent of 0.2 Mm$^2$ at $t = 51.8\,\text{min}$ when the host granule fragments in three smaller granules. At that moment, the patch of transverse magnetic flux is localized over a fragment characterised by upflow velocities of $-1.2\,\text{km s}^{-1}$, a flow likely contributing to the sheet's rise. Over the next four minutes, the area of the patch of transverse magnetic flux decreases and moves to the right edge of the new host granule (towards the direction marked by the yellow arrow). The footpoints associated with this sheet are observed at $t = 51.8\,\text{min}$, the negative-polarity one localized on an intergranular region, around $(x,y) = [3.3,3.4]$, and the positive-polarity one within the host granule. Starting from $t = 53.2\,\text{min}$, the footpoints extend around the new host granule, until it either fragments (negative-polarity footpoint) or continues to expand (positive-polarity footpoints). The magnetic flux sheet exhibits a displacement towards the direction indicated by the yellow arrow. This displacement, likely driven by plasma flows, encompasses both the sheet itself and its footpoints, which extend along the direction of displacement. Notably, the pair of footpoints appears to be oriented perpendicular to the sheet's displacement direction (towards the yellow arrow marks).

\subsection{Emerged flux} 
\label{energy}

To quantify the amount of magnetic flux carried by magnetic flux sheets, we measure the total magnetic flux emerging from each studied event and compare it with the values reported in the literature. We use the tracked longitudinal magnetic flux patches associated with the events. The patches were defined considering a threshold of three times the photon noise calculated for each dataset as described in Sect. \ref{subsec:method}. In addition, we determine by proximity which of the detected longitudinal magnetic flux patches were involved in the particular emergence event to select the corresponding footpoints, which would often be several of those. The total magnetic flux $\Phi_{\text{Lapp}}$ was calculated by averaging the magnetic flux density of the pixels within the patch and then multiplying by the area covered by the patch limited by the defined threshold. For the calculation, we used the apparent magnetic flux density output of the \texttt{sp$\_$prep.pro} routine for the Hinode SP dataset and the magnetic flux densities calculated by the weak field approximation for the SST CRISP dataset. We analyse the total magnetic flux during each event's lifetime, as well as for both associated footpoints. For events for which only one footpoint is detected, we use the maximum value reached in the observed footpoint during the event's lifetime (see examples in Sects. \ref{SG} and \ref{GL}). In the case of events for which both footpoints are observed, we noticed that most of them share similar values and evolution patterns, thus we select the maximum value reached in time and footpoint. For cases for which the values and evolution patterns are not similar, we visually determine the footpoint that best fits the evolution of the magnetic flux sheet.  

\begin{figure} 
	\centering
	\includegraphics[width=0.4\textwidth]{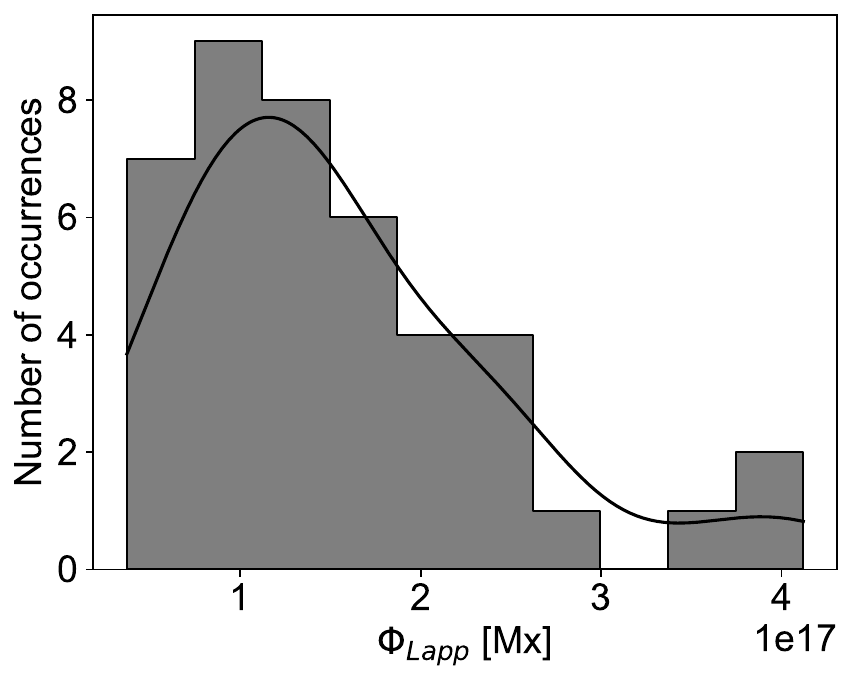}
	\caption{Histogram of the total longitudinal magnetic flux measured in the footpoints of the magnetic flux sheet events. The histogram includes the kernel density estimate providing complementary information about the shape of the distribution.}
	\label{fig:10}
\end{figure}

\begin{figure*} 
	\centering
	\includegraphics[width=\textwidth]{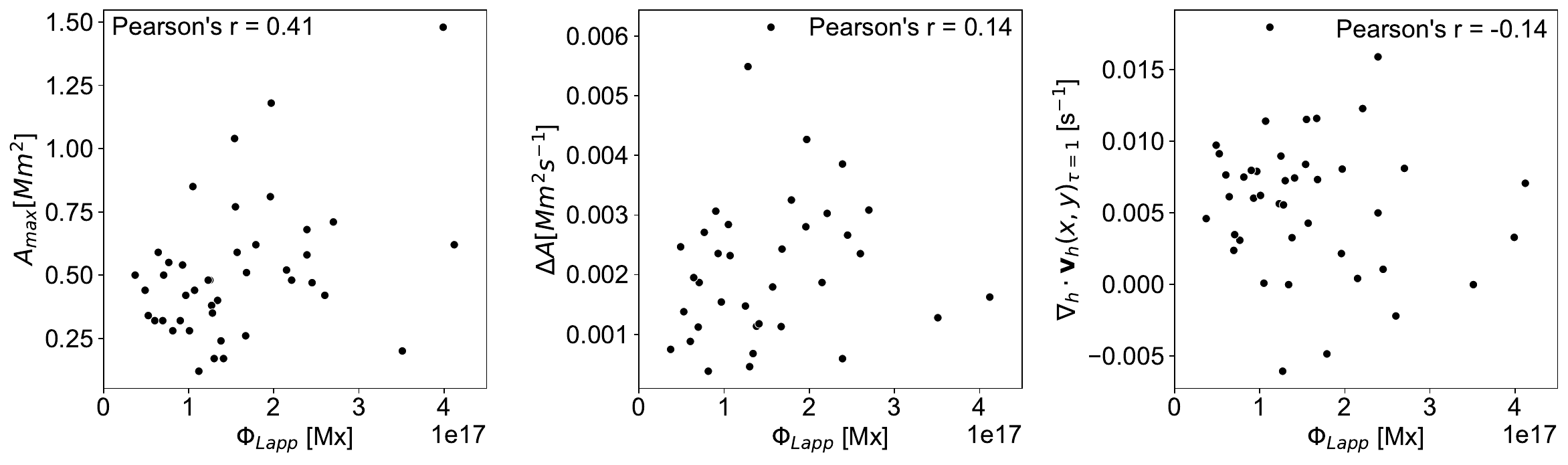}
	\caption{Scatter plots comparing the total longitudinal magnetic flux of the sheet $\Phi_{\text{Lapp}}$ to the apparent maximum area of the sheet $A_{\text{max}}$ (left panel), the rate of change of the magnetic flux sheet area $\Delta A$ (central panel), and the divergence of the horizontal flow field within the sheet $\nabla_h \cdot {\bf v}_h(x,y)_{\tau = 1}$ (right panel). The Pearson correlation coefficient is shown in each panel. See Table \ref{table:3} (Cols. 5, 6, 9) for numerical values.}
	\label{fig:11}
\end{figure*}

The histogram in Fig.\,\ref{fig:10} exhibits the distribution of total longitudinal magnetic flux $\Phi_{\text{Lapp}}$ calculated for each magnetic sheet event. The absolute values are listed in Col. 7 of Table \ref{table:3}. The average value obtained for the 42 studied events is $1.6\times10^{17}$~Mx with a slight tendency towards higher values. This average is 1.8 times larger than the mean value reported by \cite{MartinezGonzalez2009} for $\Omega$-loop type emergence and is comparable to the values of high magnetic flux events in their sample.

To understand the relationship between the emergence of magnetic flux sheets and the magnetic flux carried by their footpoints, we compare the total magnetic flux for each event with three properties of the associated magnetic flux sheet: (i) its maximum area $A_{\text{max}}$ (left panel in Fig.\,\ref{fig:11}), (ii) the rate of change of the magnetic flux sheet area $\Delta A$ (central panel in Fig.\,\ref{fig:11}) and (iii) the divergence of the horizontal flow field within the sheet measure at the surface $\nabla_h \cdot {\bf v}_h(x,y)_{\tau = 1}$ (right panel in Fig.\,\ref{fig:11}). We find that the total magnetic flux is weakly correlated with these properties, as indicated by Pearson correlation coefficients below 0.5 in all cases. In particular, the correlations between total magnetic flux with the change rate of the magnetic flux sheet area and the flow divergence are very low (correlation coefficient of 0.14), suggesting a negligible impact of total magnetic flux on these horizontal expansion metrics of the sheet. The correlation between total flux and maximum area is slightly stronger (coefficient of 0.41), but remains modest.

\subsection{Plasma dynamics} 
\label{plasma}

\begin{figure} 
	\centering
	\includegraphics[width=0.5\textwidth]{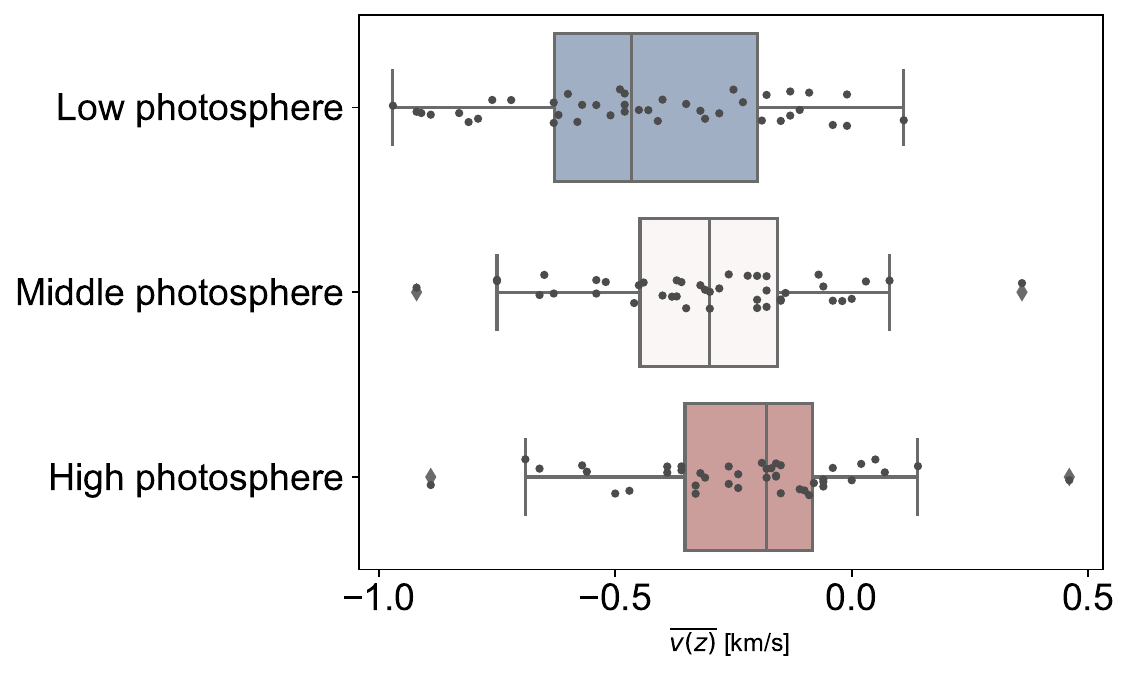}
	\caption{Average vertical velocity at different heights of the photosphere of the host granules during the emergence of their associated magnetic flux sheet. The box plots display the full distribution (scatter points), maximum and minimum values (whiskers lines), the interquartile range (box), upper and lower quartiles (box boundaries), and the median value (line within the box) for each height: low photosphere (blue box), middle photosphere (white box), and high photosphere (red box).}
	\label{fig:12}
\end{figure}

Magnetic flux sheets are coupled to the evolution of their host granule, thus the dynamics of the plasma within these granules may play a fundamental role in the occurrence and emergence process of the magnetic structure. For instance, studying the sub-photospheric origin of the magnetic flux emergence within granules, \cite{MorenoInsertis2018} found that a representative amount of the non-weakly magnetized plasma is located in upflow zones within the convection cell. In the present paper, we have analysed the vertical velocity of the granules that host flux sheets to check whether they are emerging during the lifetime of the sheet. Figure \ref{fig:12} shows a box plot of the distribution of the average of the bisector velocities at three positions of the spectral line that sample roughly three different heights in the photosphere: low (closest point to the continuum intensity), middle (roughly half maximum of the depth of the line) and high photosphere (core of the line). The box plots display the full distribution, the percentiles and the median value for each height. First, we notice that most of the distribution has negative velocities, that is, upflows. Going from the upper to the lower level, the median value becomes increasingly negative and the dispersion grows. The actual values go from $-0.18\,\text{km s}^{-1}$ in the high photosphere to $-0.45\,\text{km s}^{-1}$ in the low photosphere, confirming that most of the host granules are emerging. We identify two outlier events. The first is characterized by positive velocities in the photosphere, indicative of a host granule with a weak upflow. Inspecting the event in detail, its associated sheet emerges near an intergranular region characterized by downward flows. The second outlier involves a host granule with unusually high upward velocities (approaching $-1\,\text{km s}^{-1}$) in the middle and high photosphere, significantly exceeding the typical velocities in the sample.

Measuring the expansion rate of the magnetized elements that rise toward the photosphere, \cite{MorenoInsertis2018} found that their expansion rate at depths like $0.5$ or $1$~Mm was low enough (or negative) that they could maintain their field strength with no large weakening for several minutes. For instance, the expansion rate of the majority of elements with $B > 300$~G at depths of $0.5$ and $1$ Mm below the photosphere was below $3\times 10^{-3}$~s$^{-1}$. For upflowing plasma, at the photosphere we expect the expansion rate to be substantially larger than at depths of half or one Mm; yet, it may be of interest to compare the values obtained here with those quoted by those authors. 

\begin{figure} 
	\centering
	\includegraphics[width=0.42\textwidth]{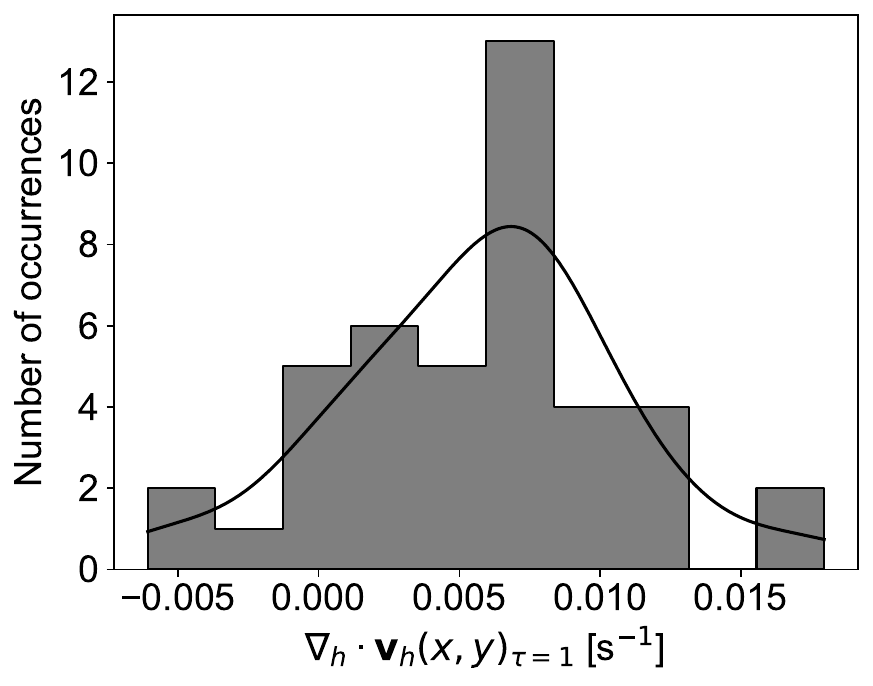}
	\caption{Histograms of the divergence of the horizontal plasma flow co-located with the magnetic flux sheet area. The horizontal velocity flow is estimated at $\tau=1$. The histogram includes the kernel density estimate providing complementary information about the shape of the distribution.}
	\label{fig:13}
\end{figure}

We analyse two potential metrics for the horizontal expansion of the magnetic flux sheet, namely the rate of change of the magnetic flux sheet area and the divergence of the horizontal flow field within the sheet measured. For the first metric, the average rate of change of the transverse magnetic flux patch area, calculated across the sample, is $1.8 \times 10^{-3}\,\text{Mm}^2 \text{s}^{-1}$ (see central panel in Fig.\,\ref{fig:11}). Assuming a linear, uniform, and circular expansion, the corresponding horizontal velocity is approximately $1.3\,\text{km s}^{-1}$. This value is comparable to the velocities of the convective plasma in granules \citep[$\sim 1\,\text{km s}^{-1}$,][]{Nordlund2009} and aligns with the joint evolution of the magnetic flux sheet and its host granule. For the second metric, we analyse the distribution of the divergence of the plasma at the location of the magnetic flux sheet while it emerges at the surface, that is, at $\tau=1$ (see right panel in Fig.\,\ref{fig:11}). Figure \ref{fig:13} shows the distribution of the mean values of the divergence of the horizontal flow field at $\tau=1$. We find that most of the values in the distribution are positive, with a distinct peak at around $ 8 \times 10^{-3}\,\text{s}^{-1}$, which provides further evidence for the emergent nature of the event. The mean value of the distribution is $ 7.5 \times 10^{-3}\,\text{s}^{-1}$ with considerable contribution of lower values. These numbers are in line with those obtained in the numerical simulation for deeper levels: the present values are larger, as expected, but still within the same order of magnitude.

\section{Summary and discussion}

We present a statistical analysis of $42$ magnetic flux sheet emergence events observed in the solar photosphere. We characterise the magnetic properties of these events and the associated plasma dynamics of their host granules. To identify these events, we developed a systematic two-step method. First, we detect patches of transverse and longitudinal magnetic flux density based on polarization signals of photospheric absorption lines. Second, we associate a single convective cell as a host granule for each event by analysing the estimated horizontal and vertical velocity fields at different photospheric heights. With our method, we confirm the distinct properties of magnetic flux sheets in the photosphere compared to subgranular-scale $\Omega$-loop type emergence \citep{MartinezGonzalez2009}. We observe that magnetic flux sheets exhibit expansive, long-lasting transverse magnetic flux patches, unlike the concentrate, nearly constant-size, and short-duration structures of $\Omega$-loop events. Our calculations of the expansion of the magnetic flux sheet, as well as the values of the divergence of the flow within the sheet and the lifetime of the transversal magnetic flux patch itself support this distinction. Additionally, we find that flux sheet footpoints often exhibit structures that expand, stretch, and fragment over time, eventually forming clusters of mixed-polarity elements. In contrast, $\Omega$-loop footpoints are typically compact, opposite-polarity patches that separate during emergence.

Our observed occurrence rates, namely $0.27$ events per day and Mm$^{2}$, align well with those predicted by the numerical simulations of \citet{MorenoInsertis2018}, namely between $0.3$ and $1$ events per day and Mm$^2$. However, given the small sample size in both the observational and numerical studies, it is difficult to draw definitive conclusions about their representativeness. Based on our study, we report the amount of magnetic flux that magnetic flux sheets contribute to the solar photosphere at an average value of $1.6\times10^{17}$ Mx per sheet (see Fig.\,\ref{fig:10}), around $1.8$ times larger than the mean value reported by \cite{MartinezGonzalez2009} for $\Omega$-loop type emergence ($0.91\times10^{17}$ Mx). Considering the estimated average magnetic flux density per sheet and appearance rate, magnetic flux sheet events could potentially contribute an average of $2.6\times10^{23}$ Mx per day to the small-scale magnetic flux that feeds the quiet Sun atmosphere. This corresponds to 24\% of the magnetic flux reported by \cite{MartinezGonzalez2009} for $\Omega$-loop type emergence and 2\% of the amount reported by \cite{Jin2009} for HIFs. We suggest that magnetic flux sheets of the type analysed in this paper could contribute a non-negligible amount of magnetic flux to the total quiet-Sun magnetic flux budget.

To delve into the process of emergence of the magnetic flux sheets, we have explored the relationship between the total longitudinal magnetic flux at their footpoints with two horizontal expansion metrics of the magnetic flux sheet, namely the rate of change of the transverse magnetic flux patch area (magnetic flux sheet area) during the emergence and the divergence of the horizontal velocity field within the sheet (see Sects. \ref{energy} and \ref{plasma}). Our findings suggest that none of these two expansion metrics are influenced by the total longitudinal flux calculated at the footpoints of the magnetic flux sheet within our sample. However, we cannot exclude the possibility that the amount of magnetic flux influences the dynamics of the three-dimensional magnetic flux sheet. Our analysis is limited to the formation heights of lines that roughly estimate different zones in the photosphere and the estimation of horizontal flows not completely in agreement with those zones, leading to two-dimensional views of the magnetic flux sheet, obscuring its full three-dimensional nature and the potential impact of the amount of magnetic flux on its dynamics. Future studies could probe the influence of magnetic flux on plasma dynamics by analysing these magnetic flux volumes at different atmospheric layers using various spectral lines.

Moreover, we investigated the relationship between the emergence of magnetic flux sheets and granular phenomena. We identified cases where the magnetic flux emerges directly coupled to nascent and rising granules, consistent with the numerical simulations of \citet{MorenoInsertis2018}. The distribution of the vertical velocity of the host granules at various photospheric heights (see Fig.\,\ref{fig:12}) confirms the emergent nature of these granules, especially in the lower photosphere. This distribution exhibits an asymmetric interquartile range, with a larger proportion of smaller velocities, potentially indicating host granules with weaker upflow velocities. While the influence of magnetic flux on convective flows is well-established in cases of strong magnetic flux \citep{Centeno2012, Wang2020}, the influence on the flows of small, subgranular features like tiny magnetic flux tubes or flux sheets is probably weak, since their plasma beta is likely to be well above unity. However, it may be of interest to investigate this point in future studies.

Additionally, our study reveals that many magnetic flux sheets emerge in association not only with nascent granules but also with exploding granules and granules with granular lanes exhibiting the diverse characteristics of emerging magnetic flux sheets.

In the case of exploding granules, the rapid expansion and fragmentation of these structures might facilitate the emergence of magnetic flux accumulated beneath the photosphere, as shown by \cite{Guglielmino2020}. Numerical simulations of exploding granules demonstrate that plasma flow in the fragmentation zone can amplify the local magnetic field a few hundred kilometres below the surface, which is later transported to the surface by shallow recirculation \citep{Rempel2018}. This could potentially contribute to magnetic flux near the fragmentation zones, in addition to the underlying emerging flux. 

In the case of granules with granular lanes, numerical simulations suggest that granular lanes can transport recycled magnetic flux from nearby intergranular lanes, regardless of the underlying magnetic flux \citep{Fischer2020}. While the significance of this recycled flux in magnetic flux emergence remains uncertain —as it could potentially occur in any granule— our findings and those of \cite{MorenoInsertis2018} offer insights. Firstly, sheets associated with granular lanes exhibit significantly lower mean horizontal flow divergence (left tail of the distribution in Fig.\,\ref{fig:13} and values in Col. 9 of Table \ref{table:3}) compared to other granular phenomena, indicating their proximity to intergranular regions. Secondly, \cite{MorenoInsertis2018} observed sub-photospheric, strongly magnetized volumes in upflows, often located near downflow regions. These observations collectively suggest that recycled magnetic flux may accompany the emergence of magnetic flux sheets.

By studying this observational sample of emerging magnetic flux sheets in the photosphere, we gain insights into the subsurface structure of emerging magnetic fields. Similar to \cite{MartinezGonzalez2009} for $\Omega$-loop type emergence, we identify cases of "double events", i.e., two events occurring in nearby granules and almost simultaneously, like, for instance, the event shown in Figure \ref{fig:5} and the events ID 58352 and ID 58366 (see Tables \ref{table:3} and \ref{table:4}). These simultaneous events may indicate a large-scale magnetic structure accumulating beneath the photosphere and emerging in sections. However, further research is needed to investigate the convective plasma conditions that facilitate the emergence of small-scale magnetic flux from the subsurface to the photosphere, particularly through numerical simulations. 

In our future work, we will analyse numerical simulations and compare their results to the observational analysis presented here. Additionally, we will focus on specific events that emerge into the chromosphere, analysing their observational characteristics and comparing them to numerical simulations.

\begin{acknowledgements}
Hinode is a Japanese mission developed and launched by ISAS/JAXA, collaborating with NAOJ as a domestic partner, NASA and STFC (UK) as international partners. Scientific operation of the Hinode mission is conducted by the Hinode science team organized at ISAS/JAXA. This team mainly consists of scientists from institutes in the partner countries. Support for the post-launch operation is provided by JAXA and NAOJ(Japan), STFC (U.K.), NASA, ESA, and NSC (Norway). The Swedish 1-m Solar Telescope is operated on the island of La Palma by the Institute for Solar Physics of Stockholm University in the Spanish Observatorio del Roque de los Muchachos of the Instituto de Astrofísica de Canarias. The Institute for Solar Physics is supported by a grant for research infrastructures of national importance from the Swedish Research Council (registration number 2021-00169). This project has received funding from the European Union’s Horizon 2020 research and innovation programme under grant agreement No 824135, the Trans-National Access Programme of SOLARNET. This research has been supported as well by the European Research Council (ERC) through the Synergy grant No. 810218 (“The Whole Sun,” ERC-2018-SyG) and by the Spanish Ministry of Science, Innovation and Universities through project PGC2018-095832-B-I00. We thank Philip Lindner and Anjali J. Kaithakkal for recording and providing us with this dataset from their observational campaign in April 2019. We thank Oleksii Andriienko for performing the data reconstruction with SSTRED. We thank Serena Criscuoli for providing the SST dataset recorded in August 2011 and Shahin Jafarzadeh and Luc Rouppe van der Voort for reducing this dataset. We thank Vigeesh Gangadharan for providing us with a time series of snapshots of his CO$^5$BOLD simulation and for his fruitful discussion and clarification regarding granular lanes. We thank Nazareth Bello-González for her comments and suggestions for improving our manuscript. S.M.D.C. and C.E.F. were funded by the Leibniz Association grant for the SAW-2018-KIS-2-QUEST project. S.L.G. acknowledge support from INAF. The National Solar Observatory is operated by the Association of Universities for Research in Astronomy, Inc. (AURA) under a cooperative agreement with the National Science Foundation. This research has made use of NASA’s Astrophysics Data System Bibliographic Services. We acknowledge the community effort devoted to the development of the following open-source packages that were used in this work: numpy (numpy.org), matplotlib (matplotlib.org), PyTorch (pytorch.org) and TensorFlow (tensorflow.org).
\end{acknowledgements}

\bibliographystyle{aa}   
\bibliography{manuscript_emstats}    

\onecolumn
\begin{appendix}

\section{Dataset details}

This section includes the characteristics of the datasets used for the present analysis. Context data information is included for completeness but is not used in this work. 

\begin{table*}[h!]
\caption{Database Hinode}              
\label{table:1}      
\centering                                      
\begin{tabular}{c c c c c p{1.7cm} p{4.5cm}}          
\hline\hline                        
Date and Time & position (x,y) & Duration& FOV & cadence& line  & Contex\\    
 & [arcsec] &  [min]   & [arcsec$^{2}$] &   [s] & [nm]  &data \\    
 \hline\hline
  2007-Sep-25 12:59UT  & [-11,6] & 180 & 2.7x41& 36&\ion{Fe}{I} 630\,nm pair & \ion{Mg}{I} 517.3\,nm FGIV  \ion{Ca}{II} H 396.8\,nm, CN 388.3\,nm FG\\ 
 \hline
 2007-Sep-26 08:15UT  & [2,6] & 111 & 2.7x41&  36&\ion{Fe}{I} 630\,nm pair & \ion{Mg}{I} 517.3\,nm FGIV \ion{Ca}{II} H 396.8\,nm, CN 388.3\,nm FG\\ 
  \hline
  2007-Sep-27 06:16UT  & [-21,6] & 37 &  2.7x41&  36&\ion{Fe}{I} 630\,nm pair & \ion{Mg}{I} 517.3\,nm FGIV \ion{Ca}{II} H 396.8\,nm, CN 388.3\,nm FG\\ 
 \hline
 2007-Sep-28 07:00UT & [-4,6] & 180  & 2.7x41 &  36&\ion{Fe}{I} 630\,nm pair &\ion{Mg}{I} 517.3\,nm FGIV \ion{Ca}{II} H 396.8\,nm, CN 388.3\,nm FG\\ 
  \hline
  2007-Sep-29 06:51UT  & [-6,6] & 173 &  2.7x41&  36&\ion{Fe}{I} 630\,nm pair &\ion{Mg}{I}  517.3\,nm FGIV \ion{Ca}{II} H 396.8\,nm, CN 388.3\,nm FG\\ 
 \hline
  2007-Oct-01 08:21UT  & [-5,7] & 108 & 2.7x41& 36&\ion{Fe}{I} 630\,nm pair &\ion{Mg}{I}  517.3\,nm FGIV \ion{Ca}{II} H 396.8\,nm, CN 388.3\,nm FG \\ 
 \hline
 2007-Oct-06 08:01UT  & [1, 6] & 137 & 2.7x41& 36&\ion{Fe}{I} 630\,nm pair &\ion{Mg}{I}  517.3\,nm FGIV\ion{Ca}{II} H 396.8\,nm, CN 388.3\,nm FG\\ 
 \hline
 2007-Oct-20 02:21UT  & [-32,6] & 99 & 4.5x41&  60&\ion{Fe}{I} 630\,nm pair & \ion{Mg}{I}  517.3\,nm FGIQUV \ion{Ca}{II} H 396.8\,nm FG \\ 
 \hline
 2007-Oct-22 18:34UT  & [10,6] & 115 & 4.5x41&  60&\ion{Fe}{I} 630\,nm pair & \ion{Mg}{I}  517.3\,nm FGIQUV  \ion{Ca}{II} H 396.8\,nm FG \\ 
  \hline
 2007-Oct-22 20:35UT  & [-188,-249] & 84 &  4.5x41&  60&\ion{Fe}{I} 630\,nm pair & \ion{Mg}{I}  517.3\,nm FGIQUV\ion{Ca}{II} H 396.8\,nm FG \\ 
 \hline
2008-Jan-16 14:01UT & [34,7] & 117  & 4.5x41 &  43&\ion{Fe}{I} 630\,nm pair &  \ion{Na}{I}  589.6\,nm FGIQUV \ion{Ca}{II} H 396.8\,nm, Gband 430.5\,nm FG \\ 
 \hline
\end{tabular}
\end{table*}

\begin{table*}[h!]
\caption{Database SST}              
\label{table:2}      
\centering                                      
\begin{tabular}{c c c c c c p{1cm} p{2.5cm}}          
\hline\hline                        
Date and Time & position (x,y) & Duration& FOV &pixelsize& cadence& line & Context\\  
   & [arcsec] &  [min]   & [arcsec$^{2}$] &  [arcsec]&  [s] &  [nm] & data\\  
 \hline\hline
 2011-Aug-06  SST1& disc center [0,0] & 47  & 57x57& 0.06 & 28 &\ion{Fe}{I} 630\,nm pair & \ion{Ca}{II}\,K narrow and broadband (imaging)\\ 
 \hline
 2019-April-29  SST2 & disc center [0,0] & 120 &50x50 & 0.059 & 28 &\ion{Fe}{I} 617\,nm, & \ion{Ca}{II} 854\,nm (spectropolarimetry) \ion{Ca}{II} K and H$\alpha$ (spectroscopy)\\
 \hline
\end{tabular}
\end{table*}

\newpage
\section{Magnetic flux sheet emergence events}

\begin{table*}[h!]
\caption{Magnetic flux sheet properties}  
\begin{center}
\label{table:3}      
\begin{tabular}{cllllllllllll}
\hline\hline
ID$^{[1]}$     & Date/Time$^{[2]}$        & Instr$^{[3]}$    & Lifetime$^{[6]}$ & A$^{[7]}$ & $\Delta$\text{A}$^{[8]}$  & $\Phi_{\text{Lapp}}$$^{[9]}$ & $\sigma_{\Phi_{\text{Lapp}}}$$^{[10]}$ & $\nabla_{h} \cdot v_{h}(x,y)_{\tau=1}$$^{[11]}$ \\
       &                   &          & [Minutes]            & [Mm$^2$]  & [Mm$^2$\,s$^{-1}$]  & [Mx] & [Mx] & [$\text{s}^{-1}$] \\
\hline\hline
100    & 20070925UT125942 & SP$^{[4]}$ & 5.4      & 0.47 & 2.66E-03    & 2.45E+17 & 1.48E+17  & 1.05E-03  \\
801    & 20070925UT125942 & SP & 5.4      & 1.18 & 4.27E-03    & 1.97E+17 & 1.33E+17  & 8.05E-03  \\
352    & 20070926UT081529 & SP & 8.4      & 0.48 & 1.47E-03    & 1.25E+17 & 7.38E+16  & 8.96E-03  \\
577    & 20070926UT081529 & SP & 6.6      & 0.51 & 2.43E-03    & 1.68E+17 & 1.15E+17  & 7.32E-03  \\
296    & 20070927UT061608 & SP & 3        & 0.52 & 1.87E-03    & 2.15E+17 & 1.01E+17  & 4.10E-04  \\
1      & 20070928UT070021 & SP & 3        & 1.48 & -$^{[12]}$   & 3.99E+17 & 3.57E+17  & 3.28E-03  \\
239    & 20070929UT065121 & SP & 7.2      & 0.85 & 2.84E-03    & 1.05E+17 & 5.07E+16  & 8.00E-05  \\
799    & 20070929UT065121 & SP & 4.8      & 0.48 & 3.03E-03    & 2.21E+17 & 1.10E+17  & 1.23E-02  \\
29     & 20071001UT082103 & SP & 8.4      & 0.32 & 8.81E-04    & 6.03E+16 & 2.63E+16  & 7.64E-03  \\
48     & 20071001UT082103 & SP & 8.4      & 0.4  & 6.78E-04    & 1.34E+17 & 7.22E+16  & -2.00E-05 \\
432    & 20071001UT082103 & SP & 4.2      & 0.55 & 2.71E-03    & 7.67E+16 & 3.68E+16  & 3.08E-03  \\
641    & 20071001UT082103 & SP & 6        & 0.5  & 1.87E-03    & 7.07E+16 & 3.52E+16  & 3.47E-03  \\
345    & 20071006UT080107 & SP & 6        & 0.5  & 7.48E-04    & 3.72E+16 & 1.65E+16  & 4.59E-03  \\
347    & 20071006UT080107 & SP & 7.8      & 0.71 & 3.08E-03    & 2.70E+17 & 1.65E+17  & 8.10E-03  \\
440    & 20071006UT080107 & SP & 7.8      & 0.58 & 5.92E-04    & 2.39E+17 & 1.36E+17  & 4.99E-03  \\
260    & 20071020UT022108 & SP & 9        & 0.28 & 3.81E-04    & 8.14E+16 & 3.62E+16  & 7.49E-03  \\
572    & 20071020UT022108 & SP & 6        & 0.62 & 3.25E-03    & 1.79E+17 & 1.58E+17  & -4.86E-03 \\
180    & 20071022UT183414 & SP & 8        & 0.42 & 1.54E-03    & 9.67E+16 & 4.04E+16  & 7.89E-03  \\
184    & 20071022UT183414 & SP & 6        & 0.54 & 2.35E-03    & 9.29E+16 & 2.75E+16  & 6.02E-03  \\
187    & 20071022UT183414 & SP & 3        & 0.44 & 2.47E-03    & 4.89E+16 & 1.76E+16  & 9.72E-03  \\
458    & 20071022UT183414 & SP & 5        & 0.32 & 1.12E-03    & 6.96E+16 & 2.56E+16  & 2.37E-03  \\
523    & 20071022UT183414 & SP & 6        & 0.59 & 1.79E-03    & 1.57E+17 & 8.48E+16  & 4.27E-03  \\
297    & 20071022UT203506 & SP & 9        & 0.59 & 1.95E-03    & 6.43E+16 & 3.10E+16  & 6.12E-03  \\
460    & 20071022UT203506 & SP & 10       & 0.81 & 2.80E-03    & 1.96E+17 & 1.20E+17  & 2.15E-03  \\
347    & 20071022UT203506 & SP & 4        & 0.28 & 0.00E+00    & 1.01E+17 & 4.82E+16  & 6.21E-03  \\
645    & 20080116UT140110 & SP & 4.3      & 1.04 & 1.22E-02    & 1.54E+17 & 8.38E+16  & 8.38E-03  \\
29372  & 20110806         & CRISP$^{[5]}$ & 7.9      & 0.24 & 1.14E-03    & 1.38E+17 & 6.62E+16  & 3.26E-03  \\
32927  & 20110806         & CRISP & 7.5      & 0.26 & 1.13E-03    & 1.67E+17 & 8.58E+16  & 1.16E-02  \\
53961  & 20110806         & CRISP & 6.1      & 0.42 & 2.35E-03    & 2.60E+17 & 1.38E+17  & -2.21E-03 \\
58570  & 20110806         & CRISP & 7        & 0.48 & -9.87E-04   & 1.23E+17 & 4.51E+16  & 5.63E-03  \\
91772  & 20110806         & CRISP & 6.5      & 0.77 & 6.15E-03    & 1.55E+17 & 7.12E+16  & 1.15E-02  \\
94010  & 20110806         & CRISP & 4.2      & 0.32 & 3.06E-03    & 9.02E+16 & 3.61E+16  & 7.96E-03  \\
12232  & 20190429         & CRISP & 4.2      & 0.17 & 4.57E-04    & 1.30E+17 & 6.66E+16  & 7.24E-03  \\
16683  & 20190429         & CRISP & 6.1      & 0.44 & 2.32E-03    & 1.07E+17 & 4.58E+16  & 1.14E-02  \\
17058  & 20190429         & CRISP & 6.5      & 0.62 & 1.62E-03    & 4.12E+17 & 1.94E+17  & 7.06E-03  \\
22153  & 20190429         & CRISP & 5.1      & 0.38 & -7.52E-04   & 1.27E+17 & 5.86E+16  & -6.07E-03 \\
39266  & 20190429         & CRISP & 9.3      & 0.2  & 1.28E-03    & 3.51E+17 & 1.85E+17  & -2.00E-05 \\
58352  & 20190429         & CRISP & 5.6      & 0.68 & 3.86E-03    & 2.39E+17 & 1.47E+17  & 1.59E-02  \\
58366  & 20190429         & CRISP & 3.7      & 0.12 & -4.90E-04   & 1.12E+17 & 5.46E+16  & 1.80E-02  \\
60914  & 20190429         & CRISP & 7        & 0.34 & 1.38E-03    & 5.26E+16 & 2.23E+16  & 9.12E-03  \\
112354 & 20190429         & CRISP & 2.8      & 0.35 & 5.49E-03    & 1.28E+17 & 5.58E+16  & 5.55E-03  \\
125941 & 20190429         & CRISP & 3.7      & 0.17 & 1.18E-03    & 1.41E+17 & 8.95E+16  & 7.43E-03  \\
\hline
\end{tabular}
\end{center}
\footnotesize{$^{[1]}$ Single and unique identifier.} \\
\footnotesize{$^{[2]}$ Date of the event detection.} \\
\footnotesize{$^{[3]}$ Instrument} \\
\footnotesize{$^{[4]}$ Hinode SP instrument.} \\
\footnotesize{$^{[5]}$ SST CRISP instrument.} \\
\footnotesize{$^{[6]}$ Event duration.} \\
\footnotesize{$^{[7]}$ Maximum area reached by the magnetic flux sheet.} \\
\footnotesize{$^{[8]}$ Change rate of the area of the magnetic flux sheet.} \\
\footnotesize{$^{[9]}$ Total magnetic flux of the magnetic flux sheet.} \\
\footnotesize{$^{[10]}$ Standard deviation of the total magnetic flux of the magnetic flux sheet.} \\
\footnotesize{$^{[11]}$ Divergence of the horizontal plasma flow at the location of the magnetic flux sheet.} \\
\footnotesize{$^{[12]}$ Event detected during its decay, thus no change was measured.} \\
\end{table*}

\newpage

\begin{table*}[h!]
\caption{Host granule properties}              
\begin{center}
\label{table:4}      
\begin{tabular}{clllllllllll}
\hline\hline
ID     & Type$^{[1]}$ & <A>$_{\text{HG}}^{[2]}$ & HG Cov$^{[3]}$ & v(z)$_{\text{low}}$$^{[4]}$ & v(z)$_{\text{middle}}$$^{[4]}$ & v(z)$_{\text{high}}$$^{[4]}$ & $\sigma_{v(z)\text{ low}}$$^{[5]}$ & $\sigma_{v(z)\text{ middle}}$$^{[5]}$ & $\sigma_{v(z)\text{ high}}$$^{[5]}$ \\
       &              & [Mm$^2$] & [\%] & [km s$^{-1}$] & [km s$^{-1}$] & [km s$^{-1}$] & [km s$^{-1}$] & [km s$^{-1}$] & [km s$^{-1}$] \\
\hline\hline
100    & GL   & 1.5          & 36          & -0.25       & -0.15          & -0.09        & 0.6             & 0.44               & 0.32             \\
801    & EG   & 1.4          & 39          & -0.72       & -0.54          & -0.39        & 0.35            & 0.32               & 0.28             \\
352    & NG   & 1.1          & 58          & -0.28       & -0.14          & -0.08        & 0.41            & 0.33               & 0.29             \\
577    & CG$^{[6]}$   &              &             &             &                &              &                 &                    &                  \\
296    & EG   & 1.1          & 31          & -0.92       & -0.75          & -0.69        & 0.42            & 0.31               & 0.24             \\
1      & EG   & 1.9          & 71          & -0.48       & -0.26          & -0.06        & 0.48            & 0.41               & 0.39             \\
239    & GL   & 1.5          & 28          & -0.6        & -0.46          & -0.36        & 0.58            & 0.45               & 0.37             \\
799    & EG   & 1.1          & 43          & -0.76       & -0.52          & -0.39        & 0.28            & 0.25               & 0.2              \\
29     & EG   & 1.6          & 20          & -0.11       & -0.06          & 0            & 0.48            & 0.36               & 0.3              \\
48     & EG   & 1.2          & 39          & -0.32       & -0.2           & -0.11        & 0.34            & 0.27               & 0.23             \\
432    & GL   & 1.5          & 20          & -0.43       & -0.2           & -0.06        & 0.39            & 0.35               & 0.32             \\
641    & GL   & 1.6          & 22          & -0.83       & -0.65          & -0.47        & 0.44            & 0.42               & 0.39             \\
345    & EG   & 1.8          & 24          & -0.91       & -0.75          & -0.66        & 0.42            & 0.32               & 0.27             \\
347    & GL   & 1.2          & 55          & -0.57       & -0.38          & -0.24        & 0.29            & 0.27               & 0.26             \\
440    & GL   & 2.3          & 20          & -0.81       & -0.66          & -0.57        & 0.51            & 0.51               & 0.48             \\
260    & CG   &              &             &             &                &              &                 &                    &                  \\
572    & NG   & 0.9          & 14          & -0.54       & -0.35          & -0.26        & 0.37            & 0.29               & 0.26             \\
180    & EG   & 3            & 14          & -0.58       & -0.37          & -0.26        & 0.35            & 0.31               & 0.29             \\
184    & CG   &              &             &             &                &              &                 &                    &                  \\
187    & EG   & 1.4          & 18          & -0.62       & -0.4           & -0.33        & 0.32            & 0.25               & 0.21             \\
458    & NG   & 0.4          & 48          & -0.09       & -0.04          & -0.04        & 0.48            & 0.37               & 0.35             \\
523    & NG   & 1.1          & 34          & -0.63       & -0.45          & -0.36        & 0.57            & 0.48               & 0.4              \\
297    & NG   & 1.2          & 30          & -0.51       & -0.28          & -0.18        & 1.22            & 0.51               & 0.42             \\
460    & NG   & 1.1          & 38          & -0.01       & 0.03           & 0.05         & 0.65            & 0.48               & 0.4              \\
347    & GL   & 1.4          & 24          & -0.01       & 0.08           & 0.14         & 0.58            & 0.54               & 0.47             \\
645    & EG   & 1.1          & 67          & -0.41       & -0.3           & -0.24        & 0.28            & 0.25               & 0.23             \\
29372  & GL   & 2.1          & 11          & -0.18       & -0.02          & 0.07         & 0.88            & 0.48               & 0.47             \\
32927  & NG   & 0.6          & 47          & -0.48       & -0.3           & -0.16        & 0.27            & 0.28               & 0.24             \\
53961  & GL   & 1.1          & 18          & -0.48       & -0.36          & -0.18        & 0.34            & 0.33               & 0.26             \\
58570  & EG   & 1.8          & 21          & 0.11        & 0.36           & 0.46         & 0.95            & 0.45               & 0.44             \\
91772  & EG   & 1.8          & 39          & -0.31       & -0.2           & -0.1         & 0.54            & 0.45               & 0.44             \\
94010  & NG   & 0.4          & 41          & -0.45       & -0.31          & -0.19        & 0.26            & 0.28               & 0.22             \\
12232  & NG   & 0.9          & 15          & -0.35       & -0.32          & -0.31        & 0.37            & 0.39               & 0.41             \\
16683  & EG   & 5            & 8           & -0.13       & -0.15          & -0.16        & 0.64            & 0.35               & 0.34             \\
17058  & EG   & 1.4          & 42          & -0.89       & -0.92          & -0.89        & 0.56            & 0.54               & 0.55             \\
22153  & NG   & 0.5          & 35          & -0.97       & -0.37          & -0.32        & 1.33            & 0.39               & 0.36             \\
39266  & GL   & 3.7          & 37          & -0.13       & -0.07          & -0.06        & 0.65            & 0.41               & 0.4              \\
58352  & NG   & 1            & 61          & -0.15       & -0.18          & -0.15        & 0.51            & 0.48               & 0.47             \\
58366  & NG   & 0.3          & 28          & -0.04       & 0              & 0.02         & 0.31            & 0.27               & 0.25             \\
60914  & CG   &              &             &             &                &              &                 &                    &                  \\
112354 & EG   & 1            & 24          & -0.49       & -0.54          & -0.56        & 0.32            & 0.32               & 0.33             \\
125941 & GL   & 1.3          & 13          & -0.23       & -0.18          & -0.15        & 0.76            & 0.44               & 0.45             \\
\hline
\end{tabular}
\end{center} 
\footnotesize{$^{[1]}$ Type of granulation: EG - Exploding granules, NG - Nascent granules, GL - Granular lanes, CG - Complex granulation.} \\
\footnotesize{$^{[2]}$ Mean area of the host granule.} \\
\footnotesize{$^{[3]}$ Percent coverage of the magnetic flux sheet over the host granule.} \\
\footnotesize{$^{[4]}$ Spatial average of the bisector velocities measured at low, middle and high heights in the photosphere.} \\
\footnotesize{$^{[5]}$ Standard deviation of the bisector velocities measured at low, middle and high heights in the photosphere.} \\
\footnotesize{$^{[6]}$ We excluded CG from the statistics on the host granule coverage and the plasma dynamics of host granules.} \\
\end{table*}

\newpage

\end{appendix} 



\end{document}